\documentclass[a4paper,fleqn]{cas-sc}

\usepackage[authoryear]{natbib}
\usepackage{graphicx}
\usepackage{amsmath}
\usepackage{amssymb}
\usepackage{amsthm}
\usepackage{amsfonts}       
\usepackage{bm}
\usepackage{multirow}
\usepackage{longtable}

\newtheorem{theorem}{Theorem}
\newtheorem{corollary}{Corollary}[theorem]

\newtheorem{remark}{Remark}
\newtheorem{definition}{Definition}
\newtheorem{lemma}{Lemma}

\def\tsc#1{\csdef{#1}{\textsc{\lowercase{#1}}\xspace}}
\tsc{WGM}
\tsc{QE}
\tsc{EP}
\tsc{PMS}
\tsc{BEC}
\tsc{DE}

\begin{document}
\let\WriteBookmarks\relax
\def\floatpagepagefraction{1}
\def\textpagefraction{.001}

\shorttitle{CV-MP: Max-Pressure Control in Heterogeneously Distributed and Partially Connected Vehicle Environments}

\shortauthors{Tan et~al.}

\title [mode = title]{\textbf{CV-MP: Max-Pressure Control in Heterogeneously Distributed and Partially Connected Vehicle Environments}}                      

\author[1,2]{ Chaopeng Tan}[orcid=0000-0003-4737-5304]
\ead{chaopeng.tan@tu-dresden.de}
\credit{Conceptualization, Methodology, Data collection, Result analysis and interpretation, Writing - Original draft preparation}

\author[1]{ Dingshan Sun}[orcid=0000-0001-9994-8181]
\ead{D.Sun-1@tudelft.nl}
\credit{Methodology, result analysis and interpretation, Writing - Original draft preparation}

\author[3]{ Hao Liu}
\ead{hao.liu@jsums.edu}
\credit{Methodology, Result analysis and interpretation, Writing - Original draft preparation}

\author[1]{ Marco Rinaldi}[orcid=0000-0001-7027-7548]
\cormark[1]
\ead{m.rinaldi@tudelft.nl}
\credit{Conceptualization, Methodology, result analysis and interpretation, Writing - Original draft preparation, Funding acquisition}

\author[1]{ Hans van Lint}[orcid=0000-0003-1493-6750]
\ead{j.w.c.vanlint@tudelft.nl}
\credit{Conceptualization, Methodology, Writing - Original draft preparation, Funding acquisition}

\cortext[cor1]{Corresponding author}

\affiliation[1]{organization={Department of Transport and Planning, Delft University of Technology},
    addressline={Gebouw 23, Stevinweg 1}, 
    city={Delft},
    postcode={2628 CN}, 
    country={The Netherlands}}

\affiliation[2]{organization={Chair of Traffic Process Automation, Technische Universität Dresden},
    addressline={Hettnerstraße 3}, 
    city={Dresden},
    postcode={01069}, 
    country={Germany}}

\affiliation[3]{organization={Department of Civil and Environmental Engineering, Jackson State University},
    addressline={1400 John R. Lynch St}, 
    city={Jackson},
    postcode={MS 39217}, 
    country={USA}}

\nonumnote{The authors would like to acknowledge the financial contribution of the EU Horizon Europe Research and Innovation Programme, Grant Agreement No. 101103808 ACUMEN.
  }

\begin{abstract}
Max-pressure (MP) control has emerged as a prominent real-time network traffic signal control strategy due to its simplicity, decentralized structure, and theoretical guarantees of network queue stability. Meanwhile, advances in connected vehicle (CV) technology have sparked extensive research into CV-based traffic signal control. Despite these developments, few studies have investigated MP control in heterogeneously distributed and partially CV environments while ensuring network queue stability. To address these research gaps, we propose a CV-based MP control (CV-MP) method that leverages real-time CV travel time information to compute the pressure, thereby incorporating both the spatial distribution and temporal delays of vehicles, unlike existing approaches that utilized only spatial distribution or temporal delays. In particular, we establish sufficient conditions for road network queue stability that are compatible with most existing MP control methods. Moreover, we pioneered the proof of network queue stability even if the vehicles are only partially connected and heterogeneously distributed, and gave a necessary condition of CV observation for maintaining the stability. Evaluation results on an Amsterdam corridor show that CV-MP significantly reduces vehicle delays compared to both actuated control and conventional MP control across various CV penetration rates. Moreover, in scenarios with dynamic traffic demand, CV-MP achieves lower spillover peaks even with low and heterogeneous CV penetration rates, further highlighting its effectiveness and robustness.
\end{abstract}




\maketitle

\section{Introduction} \label{sec: intro}

Traffic congestion remains a critical issue in urban areas, undermining mobility, economic activity, and environmental sustainability. To alleviate congestion and enhance traffic flow, effective network traffic signal control (TSC) strategies are indispensable. Architecturally, existing network traffic signal control strategies can be classified as centralized, hierarchical, and distributed control architectures. However, centralized methods, such as band-based, performance-based, and centralized MPC-based approaches \citep{yan2019network, wang2020optimizing, lin2012efficient} can be difficult to scale and adapt in the context of large, complex traffic road networks. Hierarchical methods combine high-level network optimization with localized intersection control, offering a structured framework for managing large-scale TSC \citep{sims1980sydney, gartner2001implementation, ye2016hierarchical}. Nonetheless, such multi-layered systems often struggle with partitioning, cross-level coordination \citep{rinaldi2016sensitivity}, and efficient information exchange. Distributed methods, by contrast, divide the network into smaller sub-components, i.e., intersections, allowing for localized, real-time decision-making. This localized perspective enhances scalability and reduces the communication overhead typically required by centralized or hierarchical solutions \citep{al2017distributed, wu2020distributed}.

Among distributed methods, Max-Pressure (MP) control, also referred to as Back-Pressure (BP) control, has attracted considerable attention. MP control operates effectively with local information, making it scalable and cost-effective. Its decentralized approach allows each intersection to make independent decisions, significantly reducing computational complexity and enhancing resilience to failures. Furthermore, the guarantees of queue stability and throughput optimality for MP control ensure that it can maintain efficient traffic flow within admissible demand regions \citep{varaiya2013max, varaiya2013max_1}. The basic idea of an MP control is to shift the green phase to those movements exerting the highest "pressure" on the road network, where the "pressure" is calculated as the difference in traffic state parameters between incoming and outgoing lanes weighted by saturated flow rates. The vast majority of MP controls to date have been investigating the use of different traffic state parameters for better network performance.

Early Max-Pressure (MP) control methods typically relied on aggregated traffic metrics derived from limited detection technologies. For instance, the pioneering Q-MP approach introduced by \cite{varaiya2013max} calculates the pressure of each signal phase based on vehicle counts (or queue lengths using a point-queue model). Where links have unique features (e.g., dedicated bus lanes or short links), a fixed weight can be applied to the queue length to capture these differences. However, Q-MP assumes infinite queue storage capacity, which does not hold in real-world settings. To address this issue, \cite{gregoire2014capacity} proposed Capacity-Aware MP (CA-MP), incorporating link capacity constraints. Yet, these early methods still primarily depended on vehicle count data due to the technological constraints of that era.

Recent advances in information and communication technology, particularly vehicle-to-everything (V2X) connectivity, now allow connected vehicles (CVs) to share real-time location and speed data \citep{siegel2017survey}. Unlike fixed-location detectors, which only yield aggregated traffic measurements at specific points, CVs provide two-dimensional spatiotemporal observations at the individual vehicle level, greatly enhancing traffic flow analysis and management \citep{zheng2017estimating, zhao2019various,tan2021cumulative, cao2021day, wang2024traffic}. Leveraging these rich CV data sources, researchers have proposed refined MP-based approaches. For example, \citet{li2019position} developed Position-Weighted MP (PW-MP), which assigns higher weights to queues near road inlets to mitigate spillback, and \citet{wang2022learning} proposed a learning-curve-based MP (LC-MP) that uses reinforcement learning to assign different weights to stopped and moving vehicles.

Nonetheless, most of these approaches only consider the instantaneous spatial distribution of vehicles, risking excessive delays on side roads under imbalanced flow conditions \citep{wu2018delay}. To overcome this shortcoming, some studies have explored incorporating the temporal information of vehicles for MP control. For instance, \citet{wu2018delay} introduced Head-of-Line Delay-based MP (HD-MP) to reduce delays for shorter queues, thus improving equity, and \citet{mercader2020max} implemented MP with average travel time (termed TT-MP), demonstrating favorable field results. However, neither approach explicitly demonstrated the network queue stability. Further extensions include delay-based MP (D-MP), which considers the cumulative queue length over the previous decision horizon \citep{liu2022novel}, and total delay-based MP (TD-MP), which uses total vehicle delay for pressure calculation \citep{liu2023total}; nonetheless, the network stability of TD-MP remains unproven while D-MP does not consider the spatial distribution of vehicles. 

In essence, existing MP controllers that adopt different traffic state parameters effectively assign various weights to vehicle counts. For instance, Q-MP can be regarded as applying a uniform weight of 1 to all vehicles, whereas PW-MP uses location-based weights. Notably, MP controllers that have established stability proofs typically employ time-invariant weights, relying solely on the spatial distribution of vehicles. By contrast, when time-dependent weights are introduced, as in HD-MP and TD-MP through the use of temporal information of vehicles, no formal proof of stability has been presented, despite promising experimental results.
In summary, \emph{existing MP controllers have not fully leveraged the spatiotemporal data provided by CVs, nor have they demonstrated that spatiotemporally dependent weights can achieve network queue stability.}

In addition, it is worth noting that most MP controllers leveraging vehicle-level data have only confirmed network queue stability under scenarios with 100\% CV penetration \citep{li2019position, wang2022learning, liu2022novel} or have not examined stability explicitly \citep{wu2018delay, mercader2020max, liu2023total}. Consequently, \emph{the network stability properties of MP controllers in partially CV environments are as of yet unknown.}

Moreover, in real urban networks, CVs are often heterogeneously distributed due to variations in traffic flow patterns, infrastructure availability, and socioeconomic factors \citep{goodall2013traffic, wadud2016help}. For instance, at intersections along commuter corridors, CV penetration rates may be significantly higher on main roads than on side roads during peak hours. Although numerous studies on CV-based traffic signal optimization exist \citep{feng2015real, yao2019optimization, tan2024connected, tan2024connectedcfd}, few have investigated how heterogeneously distributed CV fleets affect overall control performance. \emph{Even fewer studies have addressed how heterogeneously distributed CV penetration rates influence the stability of road network queues under CV-based MP control schemes.}

Note that while this study focuses on studying variants of MP controllers, another stream of MP control research is focused on extending MP control to a wider range of signal control applications, including signal coordination \citep{xu2024smoothing}, transit priority \citep{xu2022integrating, ahmed2024occ}, perimeter control \citep{liu2024n}, pedestrian signals \citep{liu2024max}, etc., which is outside the scope of this study and will not be discussed further.

In summary, to address the existing research gaps on variants of MP controllers, this study proposes a novel CV-based MP control (CV-MP) method that leverages CV travel time data to compute the pressure, thereby integrating both the spatial distribution and temporal delays of vehicles. The main contributions are summarized as follows:
\begin{itemize}
    \item We introduce an MP controller that leverages connected vehicle (CV) travel time data to compute pressure, fully exploiting the spatiotemporal information from CVs by incorporating both the spatial distribution and temporal delays of vehicles.
    \item We rigorously demonstrate that MP control methods with spatiotemporal-dependent weights can effectively stabilize road network queues. Additionally, we provide sufficient conditions under which MP controllers ensure queue stability, encompassing a broad range of existing MP-based approaches.
    \item We demonstrate, for the first time, that CV-MP can also stabilize traffic network queues under partial CV environments and remains effective even when CVs are heterogeneously distributed across different links. Moreover, we derive the necessary conditions of CV observations for CV-based MP controllers to ensure stability in partially CV environments.
    \item Through extensive validation, we show that CV-MP significantly outperforms both actuated control and conventional MP control methods, leading to lower vehicle delays, reduced spillovers, and improved fairness.
\end{itemize}

\section{Preliminaries} \label{sec: pre}
Before moving into the methodology, this section presents the basic network definition and traffic flow dynamics. The parameters used in this paper are summarized in Appendix~\ref{Appendix list of variables}.

\subsection{Network definition} \label{ssec: network}
\begin{figure}[!ht]
  \centering
  \includegraphics[width=0.9\textwidth]{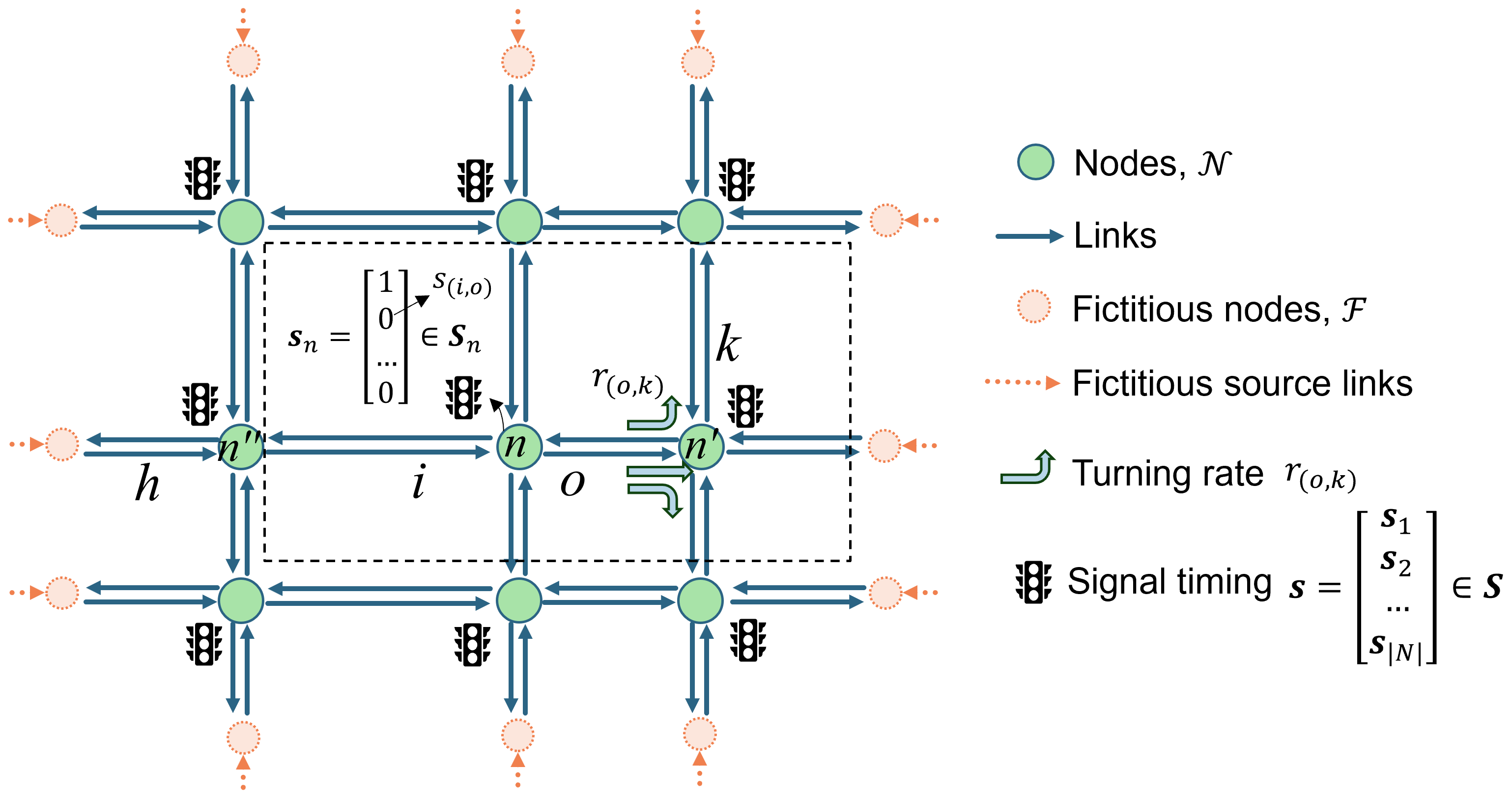}
  \caption{Network definition.}\label{fig: network}
\end{figure}

Without loss of generality, given a signalized network modeled as in Fig.~\ref{fig: network}, the notations are defined accordingly. Let $\mathcal{N}$ denote the set of nodes, i.e., signalized intersections, indexed by $n$. The set of movements for a node $n$ is denoted by $\mathcal{M}_n$, and for a neighboring node $n'$ by $\mathcal{M}_{n'}$, where each movement is indexed by a pair of incoming and outgoing links $(i,o)$ or $(o,k)$. Traffic demands are loaded from fictitious source links connecting to exogenous links by fictitious source nodes, denoted by $\mathcal{F}$. We use $\mathcal{M_N}$ and $\mathcal{M_F}$ to denote the sets of movements for all real nodes and fictitious nodes in the network, respectively. Note that fictitious source links have no physical length, thus, they are assumed to have infinite jam densities, i.e., the queue length can be accumulated infinitely. We use $\bm{s}$ to represent the overall signal vector composed of individual signal vectors $\bm{s}_n$ for each node $n$, with a binary variable $s_{(i,o)}$ being the signal state for movement $(i,o)$. Specifically, $s_{(i,o)} = 1$ represents a green phase activation state of movement $(i,o)$ and $s_{(i,o)} = 0$ otherwise. The signal state for the only movement of each fictitious node is always green. 
The set $\bm{S}$ denotes the feasible space of all signal states regarding signal phase constraints. On the incoming link of movement $(i,o)$, we use $\mathcal{J}_{(i,o)}^{cv}$ to denote the set of CVs with size $z_{(i,o)}^{cv}$ and $\mathcal{J}_{(i,o)}^{nv}$ to denote the set of non-connected vehicles (NVs) with size $z_{(i,o)}^{nv}$ and $\mathcal{J}_{(i,o)} =\mathcal{J}_{(i,o)}^{cv} \bigcup \mathcal{J}_{(i,o)}^{nv}$ to denote the set of all vehicles with size $z_{(i,o)} = z_{(i,o)}^{cv} + z_{(i,o)}^{nv}$, both of which are indexed by vehicle $j$.

\subsection{Traffic flow dynamics} \label{ssec: traffic dynamic}
Since our MP control takes spatiotemporal information at the vehicle level as input, the continuity equations are used to model the traffic flow dynamics, which is reproduced from \cite{li2019position}. For movement $(i,o) \in \mathcal{M_N}$, the traffic density at position $x$ along the incoming link $i$ that is destined to the outgoing link $o$ is defined as $\rho_{(i,o)}(x,t)$, which is a variable depending on the distance $x$ along the link and time factor $t$. While for movement $(i,o) \in \mathcal{M_F}$, the traffic density of the incoming flow is defined as $\rho_{(i,o)}(t)$, which is independent of $x$ since the fictitious source link has no physical length. Then, based on the conversation law, we have
\begin{align} 
    &\frac{\mathrm{d}\rho_{(i,o)}(t)}{\mathrm{d}t} = 
    \lambda_{(i,o)}(t)-q_{(i,o)}^{out}(s_{(i,o)}^{out}(t)), \quad \text{for} \quad (i,o) \in \mathcal{M_F}, \label{eq:dynamics-1} \\
    &\frac{\partial \rho_{(i,o)}(x,t)}{\partial t} = - \frac{\partial q_{(i,o)}(x,t)}{\partial x} \quad \text{for} \quad x\in[0,L_i], \quad (i,o) \in \mathcal{M_N}, \label{eq:dynamics-2}
\end{align} 
where $q_{(i,o)}$ denotes the flow rate at position $x$ along the incoming link $i$ that is destined to the outgoing link $o$ of movement $(i,o)$, $\lambda_{(i,o)}$ denotes the exogenous demand rate, $q_{(i,o)}^{out}$ denote the outgoing flow function of the link $i$, $s_{(i,o)}^{out}$ denote the downstream control state of the link $i$, and $L_i$ denotes the link length. In particular, for $(i,o) \in \mathcal{M_F}$, its downstream control state is always 1. 
At the boundary of the incoming link of movement $(i,o) \in \mathcal{M_N}$, we have
\begin{align}
    &\frac{\partial \rho_{(i,o)}(0,t)}{\partial t} = - \frac{\partial q_{(i,o)}(0,t)}{\partial x} = q_{(i,o)}^{in}(s_{(i,o)}^{in} (t)) - q_{(i,o)}(0,t) \quad \text{when} \quad x = 0,\label{eq:dynamics-3}\\ 
    &\frac{\partial \rho_{(i,o)}(L_i,t)}{\partial t} = - \frac{\partial q_{(i,o)}(L_i,t)}{\partial x} = q_{(i,o)}(L_i,t) - q_{(i,o)}^{out}(s_{(i,o)}^{out} (t)) \quad \text{when} \quad x = L_i,\label{eq:dynamics-4}
\end{align}
where $q_{(i,o)}^{in}$ and $s_{(i,o)}^{in}$ denote the incoming flow function and the upstream control state of the link $i$, respectively. Considering connections between links, $q_{(i,o)}^{in}(s_{(i,o)}^{in} (t))$ and $q_{(i,o)}^{out}(s_{(i,o)}^{out} (t))$ can be written as: 
\begin{align}
    &q_{(i,o)}^{out}(s_{(i,o)}^{out} (t)) = \min\{c_{(i,o)} s_{(i,o)}(t), \mu_{(i,o)}(t)\},\label{eq:dynamics-5}\\
    &q_{(i,o)}^{in}(s_{(i,o)}^{in} (t)) = \sum_{(\forall h,i)\in \mathcal{M}_{n''}} r_{(i,o)} \min\{ c_{(h,i)} s_{(h,i)}(t), \mu_{(h,i)}(t) \}, \quad \text{ if } \quad (i,o) \in \mathcal{M}_{\mathcal{N}},\label{eq:dynamics-6}
\end{align}
where $c_{(i,o)}$ is the saturated flow rate of movement $(i,o)$ at the stopline (i.e., $x = L_i$) and $\mu_{(i,o)}$ is the demand rate of movement $(i,o)$ depending on the traffic density $\rho_{(i,o)}(L_i,t)$ at the stopline. $r_{(i,o)}$ denotes the turning rate of movement $(i,o)$. 
We use $n''$ to represent the upstream neighboring node connecting to link $i$, with movements indexed by $(h,i)$. Eq.~\eqref{eq:dynamics-5} and ~\eqref{eq:dynamics-6} actually model traffic flow dynamics at the nodes, which is determined by the signal control state $s_{(i,o)}$ of movement $(i,o) \in \mathcal{M_{F \cup N}}$ at the moment $t$. Note that we do not need the exact value of $\mu_{(i,o)}$ but we know that $\mu_{(i,o)} \leq c_{(i,o)}$. This can be explained by the fact that when there are still queues for the movement, the queuing vehicles dissipate at a saturated flow rate during the green phase, making $\mu_{(i,o)} = c_{(i,o)}$. When the queue has cleared, $\mu_{(i,o)} < c_{(i,o)}$ during the remaining green phase.

\section{Methodology} \label{sec: method}
In this section, we first formulate a general form of MP controllers that use vehicle-level information, and then the CV-MP controller that uses real-time vehicle travel time information is proposed. 

\subsection{Generalized MP controller} \label{ssec: general MP}
Since existing MP controllers assume the scenario that all vehicles are connected in the modeling stage, we follow this assumption here in the generalized MP controller. 
Recall that $\mathcal{J}_{(i,o)}$ denotes the set of all vehicles of movement $(i,o) \in \mathcal{M_{F \cup N}}$ index by $j$. We use $y_j(t)$ to denote the state parameter of vehicle $j$ used for pressure calculation. Then, the generalized MP controller can be expressed as:
\begin{align}
    \bm{s}^*(t) &= \arg \max_{\bm{s} \in \bm{S}} \sum_{n \in \mathcal{N}} \left( \sum_{\forall (i,o) \in \mathcal{M}_n} s_{(i,o)}(t) c_{(i,o)} \left( \sum_{j \in \mathcal{J}_{(i,o)}} y_j(t) - \sum_{(o,\forall k) \in \mathcal{M}_{n'}} r_{(o,k)}(t) \sum_{j \in \mathcal{J}_{(o,k)}} y_j(t) \right) \right), \label{eq: general_MP}
\end{align}
where $\bm{s}^*$ denotes the signal decision by the MP controller. In essence, $\sum_{j \in \mathcal{J}_{(i,o)}} y_j(t)$ calculates the traffic state of the incoming link $i$ of movement $(i,o)$. The pressure for a movement is computed as the product of its saturated flow rate $c_{(i,o)}$ and the difference in weighted traffic states between the associated incoming and outgoing links (as indicated within the inner parentheses). Eq.~\eqref{eq: general_MP} shifts the green phase to those movements exerting the highest pressure on the road network.

\begin{figure}[ht]
    \centering
    \includegraphics[width=0.95\textwidth]{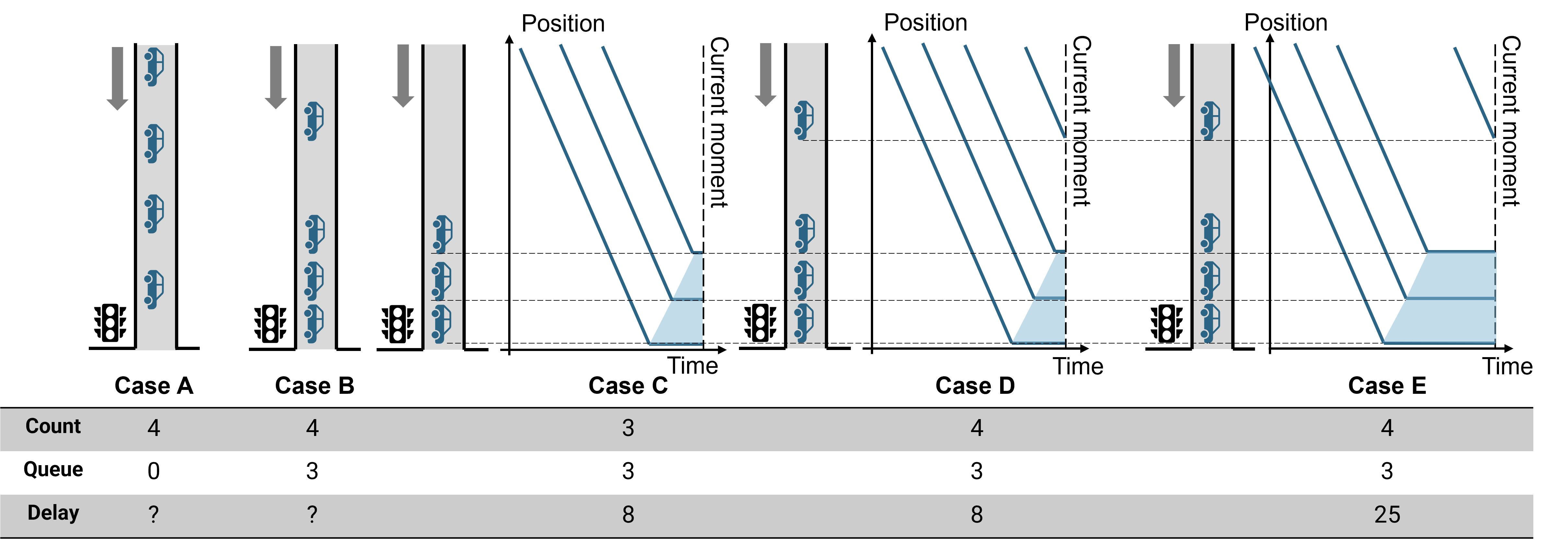}
    \caption{Different cases with the same number of vehicles}
    \label{fig: why LTT}  
\end{figure}

In the following, we will illustrate how Eq.~\eqref{eq: general_MP} encompasses existing MP controllers and elaborate on their limitations from a data utilization point of view.
Q-MP \citep{varaiya2013max} used the number of vehicles $z_{(i,o)}$ on the link for pressure calculation, i.e., letting $y_j(t) = 1$ or $y_j(t) = 1/\sqrt{L_i}$, where the latter is used if the link length $L_i$ is considered to prioritize short links that have a high probability of spillover. 
CA-MP \citep{gregoire2014capacity} adopted a similar idea, which further normalizes the queue length of the link by its capacity. 
Note that none of the above MP controllers that ignore the spatial distribution of vehicles can distinguish between the two scenarios Case A and Case B in Fig. \ref{fig: why LTT}. Obviously, Case B has a more urgent need for the green phase than Case A because of its larger queue.

Therefore, PW-MP \citep{li2019position} emphasizes queues near link inlets to mitigate spillback risks, i.e., letting $y_j(t) = x_j(t)/L_i$ if $j \in \mathcal{J}_{(i,o)}$ or $1 - x_j(t)/L_o$ if $j \in \mathcal{J}_{(o,k)}$, and $x_j$ is the distance traveled by the vehicle after entering the link.
A similar strategy was used in LC-MP \citep{wang2022learning}, which employs reinforcement learning to assign distinct position-related weights to stopped and moving vehicles for optimized performance. 
PW-MP and LC-MP can, to some extent, differentiate between Case A and B and reduce long queues on the road network by considering the spatial distribution of vehicles. However, even if the two movements have the same distribution of vehicles, it does not truly reflect their urgency to the green phase due to the potential differences in the cumulative delay incurred by these movements. As shown in \ref{fig: why LTT}, Case D and E have identical vehicle distributions, including vehicle counts and queues, but present different delays due to the different vehicle arrival times. Obviously, Case E has a more urgent need for a green phase than Case D. PW-MP and LC-MP do not distinguish between these two scenarios because they do not take into account the temporal information of the vehicles.

Considering temporal information of vehicles, HD-MP \citep{wu2018delay} used the delay $d_j(t)$ of the head-of-line vehicle for pressure calculation, i.e., letting $y_j(t) = d_j(t)$ if $j = 1$ and $0$ otherwise, which improved delay equity of vehicles. 
D-MP \citep{liu2022novel} used a delay-related indicator as the vehicle state, i.e., letting $y_j(t) = d^{T_0}_j(t)$ for all vehicles where $d^{T_0}_j$ denotes the vehicle delay during period $T_0$ between two decision steps. Essentially, $\sum_{j \in \mathcal{J}_{(i,o)}} d^{T_0}_j(t)$ is equivalent to $\sum_{t'=1}^{T_0} z_{(i,o)}^{stop}(t')$, where $z_{(i,o)}^{stop}$ denotes the number of stopped vehicles. Thus, D-MP is essentially in the same category as Q-MP and CA-MP in that it only utilizes information about the number of vehicles.
TD-MP \citep{liu2023total} modified D-MP with actual vehicle delay for pressure calculation, i.e., letting $y_j(t) = d_j(t)$ if $j \in \mathcal{J}_{(i,o)}$. HD-MP and TD-MP consider the temporal information of the vehicles to some extent; however, they ignore the spatial distribution of vehicles. That is to say, they cannot distinguish between Case C and D in Fig. \ref{fig: why LTT} as the moving vehicles are not included for pressure calculation. Obviously, Case D has a more urgent need for a green phase than Case C, though they have the same delay. In addition, considering the time information of vehicles leads to time-varying weights on link traffic density, which can cause additional difficulties in network stability proofs. Therefore, these two studies did not prove that they can stabilize road network queues.

\subsection{CV-MP}
\begin{figure}[ht]
    \centering
    \includegraphics[width=0.5\textwidth]{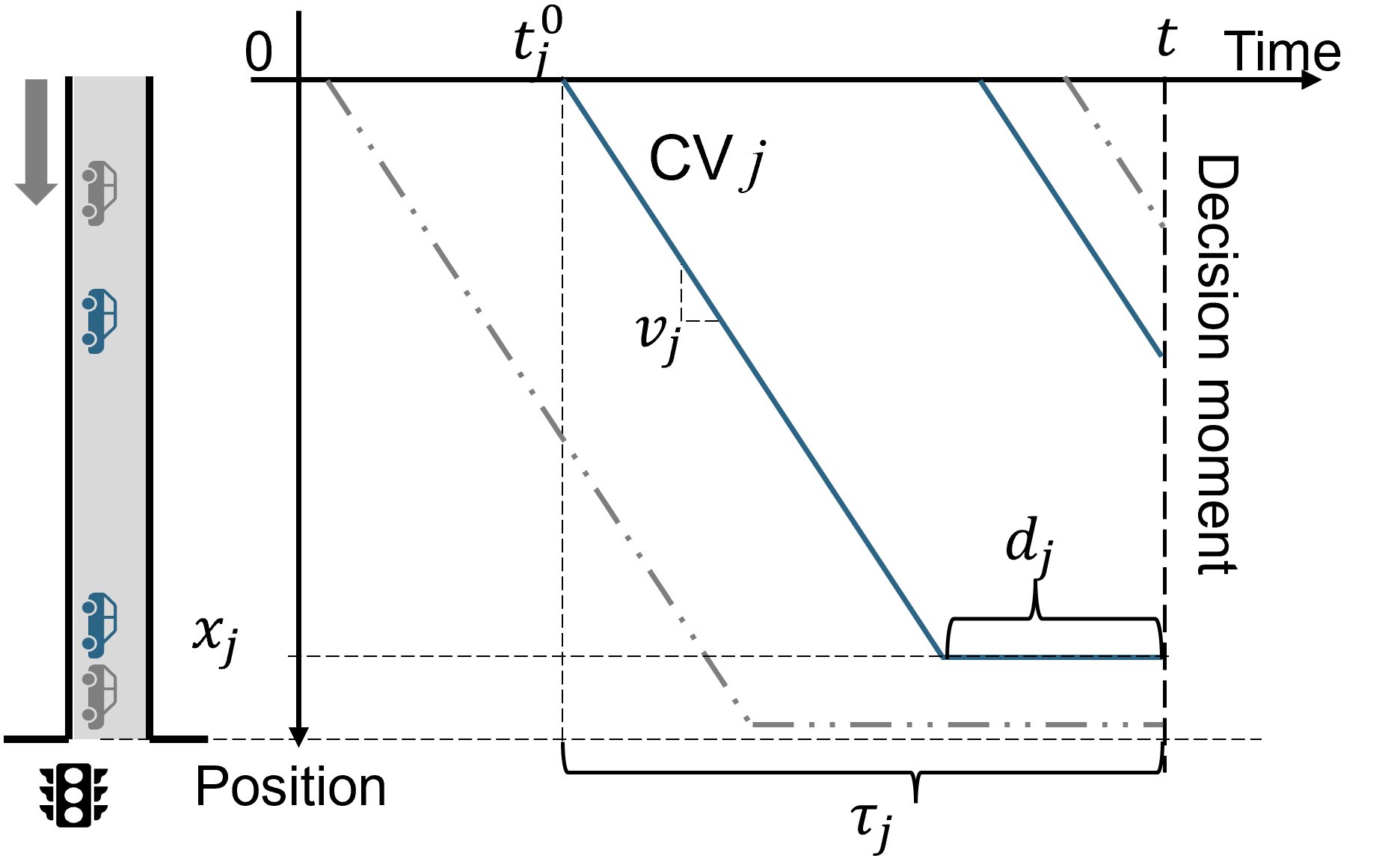}
    \caption{Travel time information of CVs}
    \label{fig: LTT}  
\end{figure}

As summarized in Section \ref{ssec: general MP}, existing studies have not fully exploited the spatiotemporal information provided by CVs, resulting in their defined pressures not adequately distinguishing between the urgent need for green phases in each movement. This study proposes to employ the link travel time information of CVs as the vehicle state parameters $y_j(t)$ to calculate the pressure in Eq.~\eqref{eq: general_MP}. Then, the CV-MP controller is written as
\begin{align}
    \textit{Practical} & \textit{ use form:} \nonumber \\
    \bm{s}^*(t) &= \arg \max_{\bm{s} \in \bm{S}} \sum_{n \in \mathcal{N}} \left( \sum_{\forall (i,o) \in \mathcal{M}_n} s_{(i,o)}(t) c_{(i,o)} \left( \sum_{j \in \mathcal{J}_{(i,o)}^{cv}} \tau_j(t) - \sum_{(o,\forall k) \in \mathcal{M}_{n'}} r_{(o,k)}(t) \sum_{j \in \mathcal{J}_{(o,k)}^{cv}} \tau_j(t) \right) \right), \label{eq: CV_MP}
\end{align}
where $\tau_j(t)$ is the normalized link travel time of CV $j$ at the decision moment $t$. As illustrated in Fig. \ref{fig: LTT}, for vehicle $j \in \mathcal{J}_{(i,o)}^{cv}$, $\tau_j$ is calculated as
\begin{align}
    \tau_j(t) = \frac{t - t_j^0}{\bar{\tau}_{(i,o)}} = \frac{1}{\bar{\tau}_{(i,o)}}\left(\frac{x_j(t)}{v_j} + d_j(t)\right), \label{eq: travel_time}
\end{align}
where $t_j^0$ is the moment the CV $j$ enters the link. $\bar{\tau}_{(i,o)}$ is the constant free-flow travel time of movement $(i,o)$ on link $i$, which can be calibrated as the link length divided by speed limits. Recall that $x_j$ is the distance traveled by the CV after entering the link and $d_j$ is the delay of the CV. The average free-flow speed of the CV is denoted by $v_j$. 
The first equal sign suggests how link travel times are obtained in practice, with CV vehicles recording their time of entry into the link $t_j^0$ and sharing real-time travel time $t - t_j^0$ to participate in the calculation of pressure.
The second equal sign divides the link travel time into free-flow travel time $x_j/v_j$ and vehicle delay $d_j$, which reveals the fact that the link travel time implies both the position information, $x_j$ employed by PW-MP and LC-MP, and the vehicle delay information, $d_j$ employed by HD-MP and TD-MP. That is to say, the proposed CV-MP in Eq.~\eqref{eq: CV_MP} exploits the spatiotemporal information of CVs.

Note that the study by \cite{mercader2020max} also uses travel time for pressure calculation, while they used the average travel time of observed vehicles collected during the last cycle, performing cyclic MP without stability proof, and this study uses real-time travel-time of CVs on the link, performing non-cyclic MP with stability proof. 

\begin{remark}[Turning intensions of CVs] \label{rmk: cv information}
    To calculate movement pressure, in addition to travel time, CVs must provide their turning intention—indicating whether they will turn left, go straight, or turn right—to classify their movement. Since many current CVs utilize map guidance services with predetermined routes, it is assumed that their turning intentions at intersections are already known for this study.
\end{remark}

Eq.~\eqref{eq: CV_MP} calculates movement pressures in the form of using CV information. By introducing binary variables $\beta_j$ to indicate whether the vehicle is connected, movement pressures can be calculated in the form of using information on all vehicles as
\begin{align}
    \textit{All vehicle} & \textit{ form:} \nonumber \\
    \bm{s}^*(t) &= \arg \max_{\bm{s} \in \bm{S}} \sum_{n \in \mathcal{N}} \left( \sum_{\forall (i,o) \in \mathcal{M}_n} s_{(i,o)}(t) c_{(i,o)} \left( \sum_{j \in \mathcal{J}_{(i,o)}} \beta_j \tau_j(t) - \sum_{(o,\forall k) \in \mathcal{M}_{n'}} r_{(o,k)}(t) \sum_{j \in \mathcal{J}_{(o,k)}} \beta_j \tau_j(t) \right) \right), \label{eq: CV_MP 1}
\end{align}
where the probability of a vehicle $j \in \mathcal{J}_{(i,o)}$ being a CV follows Bernoulli distribution, i.e., $Pr(\beta_j = 1) = \xi_{(i,o)}$. 

\begin{remark}[Heterogeneously distributed and partially CV environments] \label{rmk: h p CV}
    The proposed CV-MP does not require the CV penetration rate to be the same across the road network, as we assume a penetration rate $\xi_{(i,o)}$ for each movement $(i,o)$. That is to say, we assumed heterogeneously distributed and partially CV environments for CV-MP.
\end{remark}

We can also rewrite it in the form of traffic density and the more compact matrix form. We use $\rho_{(i,o)}^{cv}$ to denote the traffic flow density of CVs of movement $(i,o)$, then the traffic density form of CV-MP is written as:
\begin{align}
    \textit{Traffic} & \textit{ density form:} \nonumber \\
    \bm{s}^*(t) & = \arg \max_{\bm{s} \in \bm{S}} \sum_{n \in \mathcal{N}} \Big( \sum_{\forall (i,o) \in \mathcal{M}_n} s_{(i,o)}(t) c_{(i,o)} \big( \int_0^{L_i} \tau_{(i,o)}(x,t) \rho_{(i,o)}^{cv}(x,t) \mathrm{d}x  \nonumber \\
    & \qquad \qquad \qquad \qquad \qquad \qquad \qquad \qquad- \sum_{(o,\forall k) \in \mathcal{M}_{n'}} r_{(o,k)}(t) \int_0^{L_o} \tau_{(o,k)}(x,t) \rho_{(o,k)}^{cv}(x,t) \mathrm{d}x \big) \Big), \label{eq: CV_MP 2}
\end{align}
where $\tau_{(i,o)}(x,t)$ is the weight on traffic flow density of movement $(i,o)$ that is associated with vehicle travel time information $\tau_j(t)$ and 
\begin{align}
    \int_0^{L_i} \tau_{(i,o)}(x,t) \rho_{(i,o)}^{cv}(x,t) \mathrm{d}x = \sum_{j \in \mathcal{J}_{(i,o)}} \beta_j \tau_j(t) = \sum_{j \in \mathcal{J}_{(i,o)}^{cv}} \tau_j(t) \triangleq w_{(i,o)}^{cv} (t), \label{eq: CV_MP 2.1}
\end{align}
where $\triangleq$ means "defined as" and $w_{(i,o)}^{cv} (t)$ is the CV-based traffic state of movement $(i,o)$.

Obviously, $\tau_{(i,o)}(x,t)$ is spatiotemporally varying. Spatially varying means that for vehicles at different locations on the link we assign different weights to calculate the pressure. For example, in PW-MP \citep{li2019position}, $x/L_i$ or $1-x/L_i$ is used as the weight on traffic flow density at position $x$, which is spatially varying but time independent, as for the same position $x$, its weight does not change over time. Temporally varying means that the weight at position $x$ will change over time. The use of time-dependent vehicle state parameters, such as travel time and delay all result in weights that are temporally varying. For example, the travel time or delay of a stopped vehicle at position x will keep increasing until the vehicle starts moving forward. Recall that both HD-MP \citep{wu2018delay} and TD-MP \citep{liu2023total} use temporally varying weights, but they ignore the spatial distribution of vehicles. And, it remains unknown whether MP controllers considering temporally varying weights can stabilize road network queues.

For simplicity, referring to \cite{wang2022learning}, we have the matrix form of the CV-MP controller:
\begin{align}
    \textit{Matrix} & \textit{ form:} \nonumber \\
    \bm{s}^*(t) & = \arg \max_{\bm{s} \in \bm{S}} \sum_{n \in \mathcal{N}} \Big( \sum_{\forall (i,o) \in \mathcal{M}_n} s_{(i,o)}(t) c_{(i,o)} \big( w_{(i,o)}^{cv} (t) - \sum_{(o,\forall k) \in \mathcal{M}_{n'}} r_{(o,k)}(t) w_{(o,k)}^{cv} (t) \big) \Big), \nonumber \\
     & = \arg \max_{\bm{s} \in \bm{S}} (\bm{w^{cv}})^\top (\bm{I}-\bm{r})\bm{c}\bm{s}, \label{eq: CV_MP 3}
\end{align}
where $\bm{w^{cv}} \in \mathbb{R}^{|\mathcal{M_{N}}| \times 1}$ denotes the column vector of $w_{(i,o)}^{cv}(t)$ for all movements, $\bm{r} \in \mathbb{R}^{|\mathcal{M_{N}}| \times |\mathcal{M_{N}}|}$ denotes a matrix containing all turning ratios $r_{(i,o)}(t)$, $\bm{I} \in \mathbb{R}^{|\mathcal{M_{N}}| \times |\mathcal{M_{N}}|}$ is an identical matrix, and $\bm{c} \in \mathbb{R}^{|\mathcal{M_{N}}| \times |\mathcal{M_{N}}|}$ is a diagonal matrix of saturated flow rate $c_{(i,o)}$. Recall that $\bm{s} \in \mathbb{R}^{|\mathcal{M_{N}}| \times 1}$ denotes the column vector of signal state $s_{(i,o)}(t)$. 

Note that, the various forms of CV-MP controllers shown above do not consider fictitious nodes, which have only one movement and whose green signal can be considered to be active at all times. Therefore, when we consider fictitious nodes in the CV-MP controller, we only need to replace the node set $\mathcal{N}$ with $\mathcal{N} \bigcup \mathcal{F}$ and the signal decision of those real nodes in $\mathcal{N}$ does not change. In such cases, we have $w_{(i,o)}^{cv} (t) = \rho_{(i,o)}^{cv}(t)$ for $(i,o) \in \mathcal{M_{F}}$.

\section{Road Network Stability}\label{sec: stability}
In this section, we first introduce some definitions of road network stability. Then, we will provide sufficient conditions for MP controllers to stabilize the network queues, based on which, we will prove the network stability of the CV-MP controller in 100\% CV scenarios, i.e., the \emph{perfect environments}, and prove the network stability of the CV-MP controller in heterogeneously distributed and partially connected vehicle scenarios, i.e., the \emph{imperfect environments}.

\subsection{Definitions} \label{ssec: definitions}
This section introduces some definitions and lemmas for the following stability proof of MP controllers.
\subsubsection{Network stability} \label{sssec: stability}
The strong stability of the network queue length is defined as \citep{neely2022stochastic, li2019position}:
\begin{definition} [Traffic network stability] \label{def: stability}
    The traffic network queues are said to be strongly stable if the following condition holds:
    \begin{align}
        &\limsup_{T \rightarrow \infty} \frac{1}{T} \int_0^T \mathbb{E}\left( \sum_{(i,o) \in \mathcal{M_{F\cup N}}} z_{(i,o)}(t) \right) \mathrm{d}t < \infty \quad
        \text{and} \quad z_{(i,o)}(t) = \begin{cases}
            \rho_{(i,o)}(t), \text{ if } (i,o) \in \mathcal{M_F} \\
            \int_0^{L_i} \rho_{(i,o)}(x,t) \mathrm{d}x, \text{ if } (i,o) \in \mathcal{M_N}
        \end{cases} \label{eq: stability-definition}
    \end{align}
    Recall that $\rho_{(i,o)}$ denotes the traffic flow density of all vehicles of the movement $(i,o)$ and $z_{(i,o)}$ denotes the number of vehicles of the movement $(i,o)$.
\end{definition}
Definition \ref{def: stability} says that a traffic road network is strongly stable if the applied signal controller ensures that the queues on the road network do not grow infinitely over a long period of operation. As pointed out by \cite{li2019position}, there is a physical upper bound to the number of vehicles $z_{(i,o)}$ for movements $(i,o) \in \mathcal{M_{N}}$ since we consider the physical link length through the continuum equation of traffic flow dynamics. Regarding those movements $(i,o) \in \mathcal{M_{F}}$, their $z_{(i,o)}$ can be infinitely increasing, recalling that fictitious source links have no physical length. 
In other words, when the road network is unstable, congestion will be incurred on the real links with a physical upper bound of traffic density, and that congestion will keep spreading to fictitious source links, causing their queues to grow infinitely.

\begin{definition} [Lyapunov function] \label{def: lyapunov}
    Given a network with the traffic flow density state denoted as $\bm{\rho}(t)$, a Lyapunov function can be defined as
    \begin{align}
        V(\bm{\rho}(t)) \equiv \frac{1}{2} \sum_{(i,o) \in \mathcal{M}_{\mathcal{N}}} \int_0^{L_i}\int_0^{L_i} \left(y_{(i,o)}(x,t)+y_{(i,o)}(x',t)\right) \rho_{(i,o)}(x,t) \rho_{(i,o)}(x',t) \mathrm{d} x' \mathrm{d}x + \frac{1}{2} \sum_{(i,o) \in \mathcal{M}_{\mathcal{F}}}  \rho_{(i,o)}(t)^2, \label{eq: lyapunov}
    \end{align} 
    where $\rho_{(i,o)}(x,t)$ denotes the traffic flow density of all vehicles and $y_{(i,o)}(x,t)$ is a spatiotemporally varying weight on traffic flow density.
\end{definition}
In fact, $V(\bm{\rho}(t))$ is essentially the sum of the double integral of traffic flow density that is weighted $y_{(i,o)}(x,t)$, i.e., the sum of the square of the weighted $z_{(i,o)}(t)$. Reproduced from \citet{neely2022stochastic, li2019position}, a sufficient condition for traffic network stability based on the Lyapunov function can be derived. 

\begin{lemma} [Sufficient condition for traffic network stability] \label{lemma: sufficient condition}
    Given the Lyapunov function Eq. \eqref{eq: lyapunov}, if there exist constants $0<K<\infty$ and $0<\epsilon'<\infty$ such that the Lyapunov drift 
    \begin{align}
    \mathbb{E}^{\bm{\rho}(t)} \left(\frac{\mathrm{d} V(\bm{\rho}(t))}{\mathrm{d} t} \right) \leq K - \epsilon'  \mathbb{E} \left(\sum_{(i,o) \in \mathcal{M_{F\cup N}}} z_{(i,o)}(t) \right) \label{eq: stability-condition}
    \end{align}
    holds for all $t\geq0$ and all possible $\bm{\rho}(t)$, then the traffic network stability condition Eq. \eqref{eq: stability-definition} is satisfied.
\end{lemma}

\begin{proof} [Proof of Lemma \ref{lemma: sufficient condition}]
    The proof of Lemma \ref{lemma: sufficient condition} is similar to that of \cite{varaiya2013max} and \cite{li2019position}. By taking expectations and integrating the both sides of Eq.~\eqref{eq: stability-condition} over the interval $[0,T]$, we have
    \begin{align}
        \int_0^T \mathbb{E} \left(\frac{\mathrm{d} V(\bm{\rho}(t))}{\mathrm{d} t} \right) \mathrm{d}t \leq K T - \epsilon' \int_0^T \mathbb{E}\left(\sum_{(i,o) \in \mathcal{M_{F\cup N}}} z_{(i,o)}(t) \right) \mathrm{d}t, \label{eq: stability-condition proof 1}
    \end{align}
    where the superscript $\bm{\rho}(t)$ of the expectation $\mathbb{E}^{\bm{\rho}(t)}$ is dropped by taking expectations. The left side of the inequality is derived as:
    \begin{align}
        \int_0^T \mathbb{E} \left(\frac{\mathrm{d} V(\bm{\rho}(t))}{\mathrm{d} t} \right) \mathrm{d}t = \mathbb{E} \left(\int_0^T \frac{\mathrm{d} V(\bm{\rho}(t))}{\mathrm{d}t} \mathrm{d}t \right) 
         = \mathbb{E} \Big( V(\bm{\rho}(T)) \Big)- \mathbb{E} \Big( V(\bm{\rho}(0)) \Big). \label{eq: stability-condition proof 2}
    \end{align}
    Thus we have
    \begin{align}
        \mathbb{E} \Big( V(\bm{\rho}(T)) \Big)- \mathbb{E} \Big( V(\bm{\rho}(0)) \Big) \leq K T - \epsilon' \int_0^T \mathbb{E}\left(\sum_{(i,o) \in \mathcal{M_{F\cup N}}} z_{(i,o)}(t) \right) \mathrm{d}t, \label{eq: stability-condition proof 3}
    \end{align}
    Rearranging these terms and dividing them by $\epsilon' T$, we have
    \begin{align}
        \frac{1}{T} \int_0^T \mathbb{E}\left(\sum_{(i,o) \in \mathcal{M_{F\cup N}}} z_{(i,o)}(t) \right) \mathrm{d}t \leq \frac{K}{\epsilon'} + \frac{\mathbb{E} \Big( V(\bm{\rho}(0)) \Big)}{\epsilon' T} - \frac{\mathbb{E} \Big( V(\bm{\rho}(T)) \Big)}{\epsilon' T} \label{eq: stability-condition proof 4}
    \end{align}
    Since $V(\bm{\rho}(T)) \geq 0$ and $\mathbb{E} \Big( V(\bm{\rho}(0)) \Big) < \infty$, by dropping the non-positive term $- \frac{\mathbb{E} \Big( V(\bm{\rho}(T)) \Big)}{\epsilon' T}$ and taking the limit on both sides as $T \rightarrow \infty$, we have
    \begin{align}
        \limsup_{T \rightarrow \infty} \frac{1}{T} \int_0^T \mathbb{E}\left(\sum_{(i,o) \in \mathcal{M_{F\cup N}}} z_{(i,o)}(t) \right) \mathrm{d}t \leq \frac{K}{\epsilon'} + \frac{\mathbb{E} \Big( V(\bm{\rho}(0)) \Big)}{\epsilon' T} < \infty, \label{eq: stability-condition proof 5}
    \end{align}
    which proves Lemma \ref{lemma: sufficient condition}.
\end{proof}

\subsubsection{Admissible demand region} \label{sssec: demand region}
Recall that $\bm{s}$ denotes the column vector of the network signal states and $\bm{S}$ denotes the feasible polyhedron space of the network signal state. Inspired by \cite{wang2022learning}, $\bm{S}$ defines the admissible demand region $\bm{\Lambda}$ as below:
\begin{definition} [Admissible demand region] \label{def: admissible}
    Given the turning ratio $\bm{r}$ and the feasible polyhedral space $\bm{S}$ of network signal states $\bm{s}$ under certain signal constraints, the admissible demand region $\bm{\Lambda}$  of the traffic network is defined as:
    \begin{align}
        \bm{\Lambda} =\{\bm{\lambda} \mid \bm{\lambda} \preceq (\bm{I} - \bm{r})\bm{c}\bar{\bm{s}}, \exists \bar{\bm{s}} \in \bm{S}^{co}\}, \label{eq: admissible}
    \end{align}
    where $\bm{\lambda} \in \mathbb{R}^{|\mathcal{M_{F \cup N}}| \times 1}$ denotes the column vector of the average exogenous demands of the network, $\bar{\bm{s}}$ denotes the long term average of signal state $\bm{s}$, i.e., $\bar{\bm{s}} = \lim_{T \rightarrow \infty}\frac{1}{T}\sum_{t=1}^T \bm{s}$, and $\bm{S}^{co}$ is the convex hull of $\bm{S}$.
\end{definition}
Note that in Definition \ref{def: admissible}, both $\bm{I}$, $\bm{r}$, $\bm{c}$, and $\bar{\bm{s}}$ include movements at fictitious nodes, making their corresponding dimension sizes become $|\mathcal{M_{F \cup N}}|$. This admissible demand region characterizes the feasible average exogenous demand $\bm{\lambda}$ that the network can accommodate. Accordingly, a control policy is deemed throughput-optimal if it can stabilize the network queues whenever the demand lies within the interior of this region. Besides, it is proved that no controller can stabilize the network queues if the demand is located outside the admissible demand region \citep{varaiya2013max, wang2022learning}. 
Based on Eq.\eqref{eq: admissible}, we can always find a $\epsilon > 0$ for $\bm{\lambda} \in \bm{\Lambda^{int}}$ that makes 
\begin{align}
    \bm{\lambda} - (\bm{I} - \bm{r})\bm{c}\bar{\bm{s}} \preceq -\epsilon \bm{1}, \label{eq: admissible 1}
\end{align}
where $\bm{\Lambda^{int}}$ denotes the interior of $\bm{\Lambda}$ \citep{wang2022learning}.

\subsection{Sufficient conditions for MP controllers stabilizing traffic network} \label{ssec: sufficient conditions of MP}

Recall that we have used $y_{(i,o)}(x,t)$ to denote the more generalized spatiotemporally varying weight on traffic flow density in Definition \ref{def: lyapunov}. 
Then, we can rewrite the generalized MP controller presented in Eq.~\eqref{eq: general_MP} in density form as below:
\begin{align}
    \bm{s}^*(t) = \arg \max_{\bm{s} \in \bm{S}} \sum_{n \in \mathcal{N}} \Big( \sum_{\forall (i,o) \in \mathcal{M}_n} s_{(i,o)}(t) c_{(i,o)} \big( \int_0^{L_i} & y_{(i,o)}(x,t) \rho_{(i,o)}(x,t) \mathrm{d}x  \nonumber \\
    & - \sum_{(o,\forall k) \in \mathcal{M}_{n'}} r_{(o,k)}(t) \int_0^{L_o} y_{(o,k)}(x,t) \rho_{(o,k)}(x,t) \mathrm{d}x \big) \Big). \label{eq: general_MP 1}
\end{align}
Note that existing MP controllers, when all written in the density form, can be considered as variants of Eq.~\eqref{eq: general_MP 1}. For example, for Q-MP we have a fixed $y_{(i,o)}(x,t) = 1$ and for PW-MP we have a spatially varying but time-independent $y_{(i,o)}(x,t) = x/L_i$ or $1-x/L_i$ depending on whether the lane is on the incoming link or not.

\begin{remark}[Matrix form of generalized MP controller] \label{rmk: matrix generalized MP}
Include movements of fictitious nodes $\mathcal{M_{F}}$ and rewrite the generalized MP controller in Eq.~\eqref{eq: general_MP 1} into the matrix form, we have
\begin{align}
    \bm{s}^*(t) & = \arg \max_{\bm{s} \in \bm{S}} \sum_{n \in \mathcal{F \cup N}} \Big( \sum_{\forall (i,o) \in \mathcal{M}_n} s_{(i,o)}(t) c_{(i,o)} \big( w_{(i,o)}(t)   - \sum_{(o,\forall k) \in \mathcal{M}_{n'}} r_{(o,k)}(t) w_{(o,k)}(t) \big) \Big), \nonumber \\
    & = \arg \max_{\bm{s} \in \bm{S}} \bm{w}^\top (\bm{I}-\bm{r})\bm{c}\bm{s},
    \label{eq: general_MP 2} 
\end{align}
    where 
    \begin{align}
        w_{(i,o)}(t) = \begin{cases}
        \int_0^{L_i} y_{(i,o)}(x,t) \rho_{(i,o)}(x,t) \mathrm{d}x \quad \text{if } (i,o) \in \mathcal{M_{N}} \\
        \rho_{(i,o)}(t) \quad \text{if } (i,o) \in \mathcal{M_{F}}
    \end{cases}
    \end{align} and $\bm{w}$ is used to denote the column vector of $w_{(i,o)}(t)$.
\end{remark}

Before we get to the formal proof, the following lemma is needed.
Since the physical lengths of real links are considered, for a movement $(i,o) \in \mathcal{M_{N}}$, its traffic density $\rho_{(i,o)}(x,t)$ and flow rate $q_{(i,o)}(x,t)$ are naturally bounded considering the physical property of traffic flows. This is because vehicles have finite speeds and non-zero lengths, which imply finite time and space headway, leading to finite capacities and jam densities, respectively \citep{li2019position}. Without needing to know the specific values, we know that there are upper bounds, $\rho_{(i,o)}^{max}$ and $q_{(i,o)}^{max}$, of traffic density and flow rate, respectively, which makes
\begin{align}
    0 \leq \rho_{(i,o)}(x,t) \leq \rho_{(i,o)}^{max} < \infty, \label{eq: bounded density}\\ 
    0 \leq q_{(i,o)}(x,t) \leq q_{(i,o)}^{max} < \infty, \label{eq: bounded flow rate}
\end{align}
for any $x \in [0, L_i]$, any $t \geq 0$, and any $(i,o) \in \mathcal{M_{N}}$.
\begin{lemma} [Upper bounds related to traffic density and flow rate] \label{lemma: upper bounds of density and flow}
    For any $x \in [0, L_i]$, any $t \geq 0$, and any $(i,o) \in \mathcal{M_{N}}$, since the traffic density $\rho_{(i,o)}(x,t)$ and flow rate $q_{(i,o)}(x,t)$ satisfy Eq.~\eqref{eq: bounded density} and \eqref{eq: bounded flow rate}, the following inequalities hold if $0 \leq y_{(i,o)}(x,t) \leq y_{(i,o)}^{max}$:
    \begin{align}
        &\int_0^{L_i}\int_0^{L_i} \rho_{(i,o)}(x,t) \rho_{(i,o)}(x',t) \mathrm{d} x' \mathrm{d}x \leq (L_i \rho_{(i,o)}^{max})^2, \label{eq: upper bounds 1} \\
        &\int_0^{L_i} y_{(i,o)}(x,t)\rho_{(i,o)}(x,t) \mathrm{d}x \leq y_{(i,o)}^{max} L_i \rho_{(i,o)}^{max} , \label{eq: upper bounds 2} \\
        &\int_{0_{+}}^{L_{i-}} -\frac{\partial q_{(i,o)}(x',t)} {\partial x'}  \mathrm{d} x' \leq q_{(i,o)}^{max} , \label{eq: upper bounds 3} \\
        &\int_{0_{+}}^{L_{i-}} - y_{(i,o)}(x',t)\frac{\partial q_{(i,o)}(x',t)} {\partial x'}  \mathrm{d} x' \leq y_{(i,o)}^{max} L_i \dot{q}_{x,(i,o)}^{max}. \label{eq: upper bounds 4}
    \end{align}
    where $\dot{q}_{x,(i,o)}^{max}$ is a positive constant and $0 \leq \left|{\partial q_{(i,o)}(x',t)/\partial x'} \right| \leq \dot{q}_{x,(i,o)}^{max} < \infty$.
\end{lemma}
\begin{proof}[Proof of Lemma \ref{lemma: upper bounds of density and flow}]
    In Lemma \ref{lemma: upper bounds of density and flow}, Eq.~\eqref{eq: upper bounds 1} and \eqref{eq: upper bounds 2} are obvious since $y_{(i,o)} \geq 0$ and $\rho_{(i,o)}(x,t) \geq 0$. Regarding Eq.~\eqref{eq: upper bounds 3}, we have
    \begin{align}
        \int_{0_{+}}^{L_{i-}} -\frac{\partial q_{(i,o)}(x',t)} {\partial x'}  \mathrm{d} x' & \leq \left| \int_{0_{+}}^{L_{i-}} \frac{\partial q_{(i,o)}(x',t)} {\partial x'}  \mathrm{d} x' \right| = \left| q_{(i,o)}(L_{i-},t) - q_{(i,o)}(0_{+},t) \right| \nonumber \\
        & \leq \max\{q_{(i,o)}(L_{i-},t), q_{(i,o)}(0_{+},t)\} \leq q_{(i,o)}^{max},
    \end{align}
    and similarly
    \begin{align}
        \int_{0_{+}}^{L_{i-}} - y_{(i,o)}(x',t)\frac{\partial q_{(i,o)}(x',t)} {\partial x'}  \mathrm{d} x' & \leq \left| \int_{0_{+}}^{L_{i-}} y_{(i,o)}(x',t)\frac{\partial q_{(i,o)}(x',t)} {\partial x'}  \mathrm{d} x' \right| \leq y_{(i,o)}^{max} \int_{0_{+}}^{L_{i-}} \left| \frac{\partial q_{(i,o)}(x',t)} {\partial x'}\right|  \mathrm{d} x'  \nonumber \\
        & \leq y_{(i,o)}^{max} L_i \dot{q}_{x,(i,o)}^{max}.
    \end{align}
    The last inequality holds as there exists a finite positive $\dot{q}_{x,(i,o)}^{max}$ such that $\left|{\partial q_{(i,o)}(x',t)/\partial x'} \right| \leq \dot{q}_{x,(i,o)}^{max} < \infty$. Specifically, we have
    \begin{align}
        \frac{\partial q_{(i,o)}(x',t)} {\partial x'} = \frac{\mathrm{d} q_{(i,o)}(x',t)}{\mathrm{d} \rho_{(i,o)}(x',t)} \frac{\partial \rho_{(i,o)}(x',t)} {\partial x'},
    \end{align}
    where $\frac{\mathrm{d} q_{(i,o)}(x',t)}{\mathrm{d} \rho_{(i,o)}(x',t)}$ is bounded at $\rho_{(i,o)}(x',t) = 0$ or $\rho_{(i,o)}(x',t) = \rho_{(i,o)}^{max}$ according to the fundamental diagram \citep{daganzo1994cell, nishinari2014traffic} and $\frac{\partial \rho_{(i,o)}(x',t)} {\partial x'}$ is naturally bounded at traffic shock waves \citep{lighthill1955kinematic} due to physical and road limitations (e.g., driver reaction times and vehicle constraints). 
    
    Thus, Lemma \ref{lemma: upper bounds of density and flow} is proved.
\end{proof}

Regarding MP controllers in the form of Eq.~\eqref{eq: general_MP 1}, we can give sufficient conditions in Theorem \ref{thm: sufficient conditions of MP} for their ability to stabilize the road network queues. Note that \cite{wang2022learning} gave sufficient conditions of movement traffic state (i.e., $w_{(i,o)}$ in this study) for their LC-MP controller stability, while they assumed a time-independent weight on traffic flow density. In contrast, the proposed sufficient conditions in Theorem \ref{thm: sufficient conditions of MP} are more generalized.
\begin{theorem}[sufficient conditions for MP controllers stabilizing traffic network] \label{thm: sufficient conditions of MP}
    Given the admissible demand region $\bm{\Lambda}$ in Definition \ref{def: admissible}, if the exogenous demand $\bm{\lambda} \in \bm{\Lambda^{int}}$, MP controllers in the form of Eq.~\eqref{eq: general_MP 1} can strongly stabilize the traffic network queues if $y_{(i,o)}(x,t)$ where $(i,o) \in \mathcal{M_{N}}$ satisfied the following sufficient conditions: 
    \begin{enumerate}
        \item The weight $y_{(i,o)}(x,t)$ on traffic density is non-negative and upper bounded, i.e., $0 \leq y_{(i,o)}(x,t) \leq y_{(i,o)}^{max} < \infty$.
        \item $y_{(i,o)}(0,t) = 0$.
        \item $|\partial y_{(i,o)}(x,t) / \partial t|$ is upper bounded by $\Dot{y}_{(i,o)}^{max}$, i.e., $|\partial y_{(i,o)}(x,t) / \partial t| \leq \Dot{y}_{(i,o)}^{max} < \infty$.
    \end{enumerate}
\end{theorem} 

\begin{proof}[Proof of Theorem \ref{thm: sufficient conditions of MP}]
    According to Lemma \ref{lemma: sufficient condition}, the network stability under an MP controller is proved by showing that Eq. \eqref{eq: stability-condition} holds for all $t \geq 0$. First, we can decompose and simplify $\mathbb{E}^{\bm{\rho}(t)} \left(\frac{\mathrm{d} V(\bm{\rho}(t))}{\mathrm{d} t} \right)$. Based on the Leibniz integral rule and the product rule of partial differentiation, we have
    \begin{align}
        \mathbb{E}^{\bm{\rho}(t)} & \left(\frac{\mathrm{d} V(\bm{\rho}(t))}{\mathrm{d} t}\right) \nonumber \\ 
        = & \mathbb{E}^{\bm{\rho}(t)} \left(\sum_{(i,o) \in \mathcal{M}_{\mathcal{F}}} \rho_{(i,o)}(t)\frac{\mathrm{d} \rho_{(i,o)}(t)}{\mathrm{d} t} \right) + \mathbb{E}^{\bm{\rho}(t)} \left( \sum_{(i,o) \in \mathcal{M}_{\mathcal{N}}} \int_0^{L_i}\int_0^{L_i} \frac{\partial y_{(i,o)}(x,t)}{\partial t} \rho_{(i,o)}(x,t) \rho_{(i,o)}(x',t) \mathrm{d} x' \mathrm{d}x \right) \nonumber \\
        & + \mathbb{E}^{\bm{\rho}(t)} \left( \sum_{(i,o) \in \mathcal{M}_{\mathcal{N}}} \int_0^{L_i}\int_0^{L_i} \left(y_{(i,o)}(x,t)+y_{(i,o)}(x',t)\right) \left(\rho_{(i,o)}(x,t)\frac{\partial \rho_{(i,o)}(x',t)} {\partial t} \right) \mathrm{d} x' \mathrm{d}x \right) \nonumber \\ 
        = & -\mathbb{E}^{\bm{\rho}(t)} \left(\sum_{(i,o) \in \mathcal{M}_{\mathcal{F}}} \rho_{(i,o)}(t)\frac{\mathrm{d} q_{(i,o)}(t)}{\mathrm{d} x} \right) + \underbrace{\mathbb{E}^{\bm{\rho}(t)} \left( \sum_{(i,o) \in \mathcal{M}_{\mathcal{N}}} \int_0^{L_i}\int_0^{L_i} \frac{\partial y_{(i,o)}(x,t)}{\partial t} \rho_{(i,o)}(x,t) \rho_{(i,o)}(x',t) \mathrm{d} x' \mathrm{d}x \right)}_{\delta_0} \nonumber \\
        & - \mathbb{E}^{\bm{\rho}(t)} \left( \sum_{(i,o) \in \mathcal{M}_{\mathcal{N}}} \int_0^{L_i}\int_0^{L_i} \left(y_{(i,o)}(x,t)+y_{(i,o)}(x',t)\right) \left(\rho_{(i,o)}(x,t)\frac{\partial q_{(i,o)}(x',t)} {\partial x'} \right) \mathrm{d} x' \mathrm{d}x \right), \label{eq: sufficient conditions proof 0}
    \end{align}
    where the second equal sign is obtained by replacing ${\partial \rho_{(i,o)}(t)}/{\partial t}$ and ${\partial \rho_{(i,o)}(x',t)}/{\partial t}$ with $-{\partial q_{(i,o)}(t)}/{\partial x}$ and $-{\partial q_{(i,o)}(x',t)}/{\partial x'}$ according to Eq.~\eqref{eq:dynamics-1} and \eqref{eq:dynamics-2}.
    Referring to \cite{li2019position}, the third term in Eq.~\eqref{eq: sufficient conditions proof 0} can be further decomposed by substituting $x'=0$ and $x'=L_i$ to consider boundary dynamics:
    \begin{align}
        - &\mathbb{E}^{\bm{\rho}(t)} \left( \sum_{(i,o) \in \mathcal{M}_{\mathcal{N}}} \int_0^{L_i}\int_0^{L_i} \left(y_{(i,o)}(x,t)+y_{(i,o)}(x',t)\right) \left(\rho_{(i,o)}(x,t)\frac{\partial q_{(i,o)}(x',t)} {\partial x'} \right) \mathrm{d} x' \mathrm{d}x \right) \nonumber \\
        = & - \mathbb{E}^{\bm{\rho}(t)} \left( \sum_{(i,o) \in \mathcal{M}_{\mathcal{N}}} \frac{\partial q_{(i,o)}(0,t)} {\partial x'} \int_0^{L_i} \left(y_{(i,o)}(x,t)+y_{(i,o)}(0,t)\right) \rho_{(i,o)}(x,t) \mathrm{d}x \right) \nonumber \\
        & - \mathbb{E}^{\bm{\rho}(t)} \left( \sum_{(i,o) \in \mathcal{M}_{\mathcal{N}}} \frac{\partial q_{(i,o)}(L_i,t)} {\partial x'} \int_0^{L_i} \left(y_{(i,o)}(x,t)+y_{(i,o)}(L_i,t)\right) \rho_{(i,o)}(x,t) \mathrm{d}x \right) \nonumber \\ 
        & - \mathbb{E}^{\bm{\rho}(t)} \left( \sum_{(i,o) \in \mathcal{M}_{\mathcal{N}}} \int_0^{L_i}\int_{0_{+}}^{L_{i-}} \left(y_{(i,o)}(x,t)+y_{(i,o)}(x',t)\right) \left(\rho_{(i,o)}(x,t)\frac{\partial q_{(i,o)}(x',t)} {\partial x'} \right) \mathrm{d} x' \mathrm{d}x \right) \nonumber \\
        =& \underbrace{\sum_{(i,o) \in \mathcal{M}_{\mathcal{N}}} \mathbb{E}^{\bm{\rho}(t)} \left( q_{(i,o)}^{in}(s_{(i,o)}^{in}(t)) \int_0^{L_i} y_{(i,o)}(x,t) \rho_{(i,o)}(x,t) \mathrm{d}x \right) }_{\delta_1} - \underbrace{\sum_{(i,o) \in \mathcal{M}_{\mathcal{N}}} \mathbb{E}^{\bm{\rho}(t)} \left( q_{(i,o)}(0,t) \int_0^{L_i} y_{(i,o)}(x,t) \rho_{(i,o)}(x,t) \mathrm{d}x \right) }_{\delta_4} \nonumber \\
        & + \underbrace{\sum_{(i,o) \in \mathcal{M}_{\mathcal{N}}} \mathbb{E}^{\bm{\rho}(t)} \left( \left(q_{(i,o)}^{in}(s_{(i,o)}^{in}(t))-q_{(i,o)}(0,t)\right) \int_0^{L_i} y_{(i,o)}(0,t) \rho_{(i,o)}(x,t) \mathrm{d}x \right) }_{=0} \nonumber \\
        & - \underbrace{\sum_{(i,o) \in \mathcal{M}_{\mathcal{N}}} \mathbb{E}^{\bm{\rho}(t)} \left( q_{(i,o)}^{out}(s_{(i,o)}^{out}(t)) \int_0^{L_i} y_{(i,o)}(x,t) \rho_{(i,o)}(x,t) \mathrm{d}x \right) }_{\delta_2} \nonumber \\
        & - \underbrace{\sum_{(i,o) \in \mathcal{M}_{\mathcal{N}}} \mathbb{E}^{\bm{\rho}(t)} \left( q_{(i,o)}^{out}(s_{(i,o)}^{out}(t)) \int_0^{L_i}y_{(i,o)}(L_i,t) \rho_{(i,o)}(x,t) \mathrm{d}x \right) }_{\delta_3} \nonumber \\
        & + \underbrace{\sum_{(i,o) \in \mathcal{M}_{\mathcal{N}}} \mathbb{E}^{\bm{\rho}(t)} \left( q_{(i,o)}(L_i,t) \int_0^{L_i} \left(y_{(i,o)}(x,t) + y_{(i,o)}(L_i,t)\right) \rho_{(i,o)}(x,t) \mathrm{d}x \right) }_{\delta_5} \nonumber \\
        & - \underbrace{\sum_{(i,o) \in \mathcal{M}_{\mathcal{N}}} \mathbb{E}^{\bm{\rho}(t)} \left( \int_0^{L_i} \int_{0_{+}}^{L_{i-}} \left(y_{(i,o)}(x,t)+y_{(i,o)}(x',t)\right) \left(\rho_{(i,o)}(x,t) \frac{\partial q_{(i,o)}(x',t)} {\partial x'} \right) \mathrm{d} x' \mathrm{d}x \right) }_{\delta_6}, \label{eq: sufficient conditions proof 0.1}
    \end{align}
    where the second equal sign is obtained by substituting Eq.~\eqref{eq:dynamics-3} and \eqref{eq:dynamics-4} to introduce the signal control policy for traffic network stability. The third term in Eq.~\eqref{eq: sufficient conditions proof 0.1} with $y_{(i,o)}(0,t)$ can be dropped as $y_{(i,o)}(0,t) = 0$ based on \emph{Condition 2} in Theorem \ref{thm: sufficient conditions of MP}. 
    Similarly, the first term in Eq.~\eqref{eq: sufficient conditions proof 0} is also decomposed by substituting Eq.~\eqref{eq:dynamics-1}:
    \begin{align}
        -\mathbb{E}^{\bm{\rho}(t)} \left(\sum_{(i,o) \in \mathcal{M}_{\mathcal{F}}} \rho_{(i,o)}(t)\frac{\mathrm{d} q_{(i,o)}(t)}{\mathrm{d} x} \right) = \underbrace{\sum_{(i,o) \in \mathcal{M}_{\mathcal{F}}} \mathbb{E}^{\bm{\rho}(t)} \left( \left( \lambda_{(i,o)}(t)-q_{(i,o)}^{out}(s_{(i,o)}^{out}(t)) \right)\rho_{(i,o)}(t) \right)}_{\eta}. \label{eq: sufficient conditions proof 0.2}
    \end{align}
    
    Combining Eq.~\eqref{eq: sufficient conditions proof 0}-\eqref{eq: sufficient conditions proof 0.2}, we have
    \begin{align}
        \mathbb{E}^{\bm{\rho}(t)} & \left(\frac{\mathrm{d} V(\bm{\rho}(t))}{\mathrm{d} t}\right) =\eta + \delta_0 + \delta_1 - \delta_2 - \delta_3 - \delta_4 + \delta_5 - \delta_6. \label{eq: sufficient conditions proof 1}
    \end{align}
    
    Specifically, we have
    \begin{align}
        \delta_0 & = \sum_{(i,o) \in \mathcal{M}_{\mathcal{N}}} \mathbb{E}^{\bm{\rho}(t)} \left(\int_0^{L_i}\int_0^{L_i} \frac{\partial y_{(i,o)}(x,t)}{\partial t} \rho_{(i,o)}(x,t) \rho_{(i,o)}(x',t) \mathrm{d} x' \mathrm{d}x \right) \nonumber \\
        & \leq \sum_{(i,o) \in \mathcal{M}_{\mathcal{N}}} \mathbb{E}^{\bm{\rho}(t)} \left(\int_0^{L_i}\int_0^{L_i} \left|\frac{\partial y_{(i,o)}(x,t)}{\partial t}\right| \rho_{(i,o)}(x,t) \rho_{(i,o)}(x',t) \mathrm{d} x' \mathrm{d}x \right) \nonumber \\
        & \leq \sum_{(i,o) \in \mathcal{M}_{\mathcal{N}}} \mathbb{E}^{\bm{\rho}(t)} \left(\int_0^{L_i}\int_0^{L_i} \Dot{y}_{(i,o)}^{max} \rho_{(i,o)}(x,t) \rho_{(i,o)}(x',t) \mathrm{d} x' \mathrm{d}x \right) \nonumber \\
        & \leq \sum_{(i,o) \in \mathcal{M}_{\mathcal{N}}} \Dot{y}_{(i,o)}^{max} (L_i \rho_{(i,o)}^{max})^2  = K_0. \label{eq: ub of delta_0}
    \end{align}
    The second less-than-and-equal sign is based on \emph{Condition 3} of Theorem \ref{thm: sufficient conditions of MP}, and the third less-than-and-equal sign is based on Lemma \ref{lemma: upper bounds of density and flow}.
    \begin{align}
        - \delta_3 = -\sum_{(i,o) \in \mathcal{M}_{\mathcal{N}}} \mathbb{E}^{\bm{\rho}(t)} \left( q_{(i,o)}^{out}(s_{(i,o)}^{out}(t)) \int_0^{L_i} y_{(i,o)}(L_i,t) \rho_{(i,o)}(x,t) \mathrm{d}x \right) \leq 0, \label{eq: ub of delta_3}
    \end{align}
    since $q_{(i,o)}^{out}(s_{(i,o)}^{out}(t)) \geq 0$ based on Eq.~\eqref{eq:dynamics-5}, $y_{(i,o)}(L_i,t) \geq 0$ based on \emph{Condition 1} of Theorem \ref{thm: sufficient conditions of MP}, and $\rho_{(i,o)}(x,t) \geq 0$ based on Eq.~\eqref{eq: bounded density}. Similarly, as $q_{(i,o)}(0,t) \geq 0$ based on Eq.~\eqref{eq: bounded flow rate}, we have
    \begin{align}
        - \delta_4 = - \sum_{(i,o) \in \mathcal{M}_{\mathcal{N}}} \mathbb{E}^{\bm{\rho}(t)} \left( q_{(i,o)}(0,t) \int_0^{L_i} y_{(i,o)}(x,t) \rho_{(i,o)}(x,t) \mathrm{d}x \right) \leq 0. \label{eq: ub of delta_4}
    \end{align}
    Further, 
    \begin{align}
        \delta_5 & = \sum_{(i,o) \in \mathcal{M}_{\mathcal{N}}} \mathbb{E}^{\bm{\rho}(t)} \left( q_{(i,o)}(L_i,t) \int_0^{L_i} \left(y_{(i,o)}(x,t) + y_{(i,o)}(L_i,t)\right) \rho_{(i,o)}(x,t) \mathrm{d}x \right) \nonumber \\
        & \leq \sum_{(i,o) \in \mathcal{M}_{\mathcal{N}}} 2 q_{(i,o)}^{max} y_{(i,o)}^{max} L_i \rho_{(i,o)}^{max}  = K_1, \label{eq: ub of delta_5}
    \end{align}
    where the less-than-and-equal sign is based on Eq.~\eqref{eq: bounded flow rate} and Lemma \ref{lemma: upper bounds of density and flow}.
    \begin{align}
        -\delta_6 & = - \sum_{(i,o) \in \mathcal{M}_{\mathcal{N}}} \mathbb{E}^{\bm{\rho}(t)} \left( \int_0^{L_i} \int_{0_{+}}^{L_{i-}} \left(y_{(i,o)}(x,t)+y_{(i,o)}(x',t)\right) \left(\rho_{(i,o)}(x,t) \frac{\partial q_{(i,o)}(x',t)} {\partial x'} \right) \mathrm{d} x' \mathrm{d}x \right) \nonumber \\
        & = \sum_{(i,o) \in \mathcal{M}_{\mathcal{N}}} \mathbb{E}^{\bm{\rho}(t)} \left( \int_0^{L_i} \int_{0_{+}}^{L_{i-}} -y_{(i,o)}(x,t) \rho_{(i,o)}(x,t) \frac{\partial q_{(i,o)}(x',t)}{\partial x'} -\rho_{(i,o)}(x,t) y_{(i,o)}(x',t)\frac{\partial q_{(i,o)}(x',t)} {\partial x'}  \mathrm{d} x' \mathrm{d}x \right) \nonumber \\
        & = \sum_{(i,o) \in \mathcal{M}_{\mathcal{N}}} \mathbb{E}^{\bm{\rho}(t)} \left( \int_0^{L_i}  y_{(i,o)}(x,t) \rho_{(i,o)}(x,t) \left( \int_{0_{+}}^{L_{i-}} -\frac{\partial q_{(i,o)}(x',t)}{\partial x'} \mathrm{d} x'\right) \mathrm{d}x \right) \nonumber \\
        & \qquad + \sum_{(i,o) \in \mathcal{M}_{\mathcal{N}}} \mathbb{E}^{\bm{\rho}(t)} \left(\int_0^{L_i} \rho_{(i,o)}(x,t) \left(\int_{0_{+}}^{L_{i-}} - y_{(i,o)}(x',t)\frac{\partial q_{(i,o)}(x',t)} {\partial x'}  \mathrm{d} x'\right) \mathrm{d}x \right) \nonumber \\
        & \leq \sum_{(i,o) \in \mathcal{M}_{\mathcal{N}}} \mathbb{E}^{\bm{\rho}(t)} \left( q_{(i,o)}^{max} \int_0^{L_i}  y_{(i,o)}(x,t) \rho_{(i,o)}(x,t) \mathrm{d}x \right) + \sum_{(i,o) \in \mathcal{M}_{\mathcal{N}}} \mathbb{E}^{\bm{\rho}(t)} \left(y_{(i,o)}^{max} L_i \dot{q}_{x,(i,o)}^{max} \int_0^{L_i} \rho_{(i,o)}(x,t) \mathrm{d}x \right) \nonumber \\
        & \leq \sum_{(i,o) \in \mathcal{M}_{\mathcal{N}}} q_{(i,o)}^{max} y_{(i,o)}^{max} L_i \rho_{(i,o)}^{max} + y_{(i,o)}^{max} L_i^2 \dot{q}_{x,(i,o)}^{max} \rho_{(i,o)}^{max} = K'_1, \label{eq: ub of delta_6}
    \end{align}
    where the second equal sign is obtained by separating the term $\left(y_{(i,o)}(x,t)+y_{(i,o)}(x',t)\right)$, the third equal sign is obtained by separating integrals over $x'$ and over $x$, and the inequalities are obtained based on Eq.~\eqref{eq: upper bounds 2}-\eqref{eq: upper bounds 4} in Lemma \ref{lemma: upper bounds of density and flow}.

    The remaining terms $\eta$, $\delta_1$, and $\delta_2$ are all associated with the signal control policy, i.e., $q_{(i,o)}^{in}(s_{(i,o)}^{in}(t))$ and $q_{(i,o)}^{out}(s_{(i,o)}^{out}(t))$. Then, we have
    \begin{align}
        \eta + \delta_1 - \delta_2 = & \sum_{(i,o) \in \mathcal{M}_{\mathcal{F}}} \mathbb{E}^{\bm{\rho}(t)} \left( \left( \lambda_{(i,o)}(t)-q_{(i,o)}^{out}(s_{(i,o)}^{out}(t)) \right)\rho_{(i,o)}(t) \right) \nonumber \\
        & + \sum_{(i,o) \in \mathcal{M}_{\mathcal{N}}} \mathbb{E}^{\bm{\rho}(t)} \left( q_{(i,o)}^{in}(s_{(i,o)}^{in}(t)) \int_0^{L_i} y_{(i,o)}(x,t) \rho_{(i,o)}(x,t) \mathrm{d}x \right) \nonumber \\
        & - \sum_{(i,o) \in \mathcal{M}_{\mathcal{N}}} \mathbb{E}^{\bm{\rho}(t)} \left( q_{(i,o)}^{out}(s_{(i,o)}^{out}(t)) \int_0^{L_i} y_{(i,o)}(x,t) \rho_{(i,o)}(x,t) \mathrm{d}x \right) \nonumber \\
        = & \mathbb{E}^{\bm{\rho}(t)} \left(\sum_{(i,o) \in \mathcal{M}_{\mathcal{F}}} \left( \left(\lambda_{(i,o)}(t) -q_{(i,o)}^{out}(s_{(i,o)}^{out}(t))   \right)w_{(i,o)}(t) \right) \right) \nonumber \\
        &+  \mathbb{E}^{\bm{\rho}(t)} \left( \sum_{(i,o) \in \mathcal{M}_{\mathcal{N}}} \left( q_{(i,o)}^{in}(s_{(i,o)}^{in}(t)) - q_{(i,o)}^{out}(s_{(i,o)}^{out}(t)) \right) w_{(i,o)}(t) \right). \label{eq: eta_delta12 1}
    \end{align}
    where the second equal sign is obtained by replacing $\rho_{(i,o)}(t)$ and $\int_0^{L_i} y_{(i,o)}(x,t) \rho_{(i,o)}(x,t) \mathrm{d}x$ with $w_{(i,o)}(t)$. Substituting $q_{(i,o)}^{in}(s_{(i,o)}^{in}(t))$ and $q_{(i,o)}^{out}(s_{(i,o)}^{out}(t))$ based on Eq. \eqref{eq:dynamics-5} and \eqref{eq:dynamics-6}, rearranging terms, and rewriting $\eta + \delta_1 - \delta_2$ in matrix form, we have
    \begin{align}
        \eta + \delta_1 - \delta_2 & =  \mathbb{E}^{\bm{\rho}(t)} \left(\sum_{(i,o) \in \mathcal{M}_{\mathcal{F}}} \left( \left(\lambda_{(i,o)}(t) -\min\{c_{(i,o)} s_{(i,o)}(t), \mu_{(i,o)}(t)\} )   \right)w_{(i,o)}(t) \right) \right) \nonumber \\
        +  \mathbb{E}^{\bm{\rho}(t)} & \left( \sum_{(i,o) \in \mathcal{M}_{\mathcal{N}}} \left( \sum_{(\forall h,i)\in \mathcal{M}_{n''}} r_{(i,o)} \min\{ c_{(h,i)} s_{(h,i)}(t), \mu_{(h,i)}(t) \}- \min\{c_{(i,o)} s_{(i,o)}(t), \mu_{(i,o)}(t)\} \right) w_{(i,o)}(t) \right)\nonumber \\
        & = \mathbb{E}^{\bm{\rho}(t)} \left( \bm{w}^{\top} \left( \bm{\lambda} - (\bm{I} - \bm{r}) \min\{\bm{\mu}, \bm{c}\bm{s}^*\} \right) \right) \nonumber \\
        & = \mathbb{E}^{\bm{\rho}(t)} \left( \bm{w}^{\top} \left( \bm{\lambda} - (\bm{I} - \bm{r})\bm{c}\bm{s}^* + (\bm{I} - \bm{r})\bm{c}\bm{s}^* - (\bm{I} - \bm{r}) \min\{\bm{\mu}, \bm{c}\bm{s}^*\} \right) \right) \nonumber \\ 
        & = \underbrace{\mathbb{E}^{\bm{\rho}(t)} \left( \bm{w}^{\top} \left( \bm{\lambda} - (\bm{I} - \bm{r})\bm{c}\bm{s}^* \right) \right)}_{\eta_1} + 
        \underbrace{\mathbb{E}^{\bm{\rho}(t)} \left( \bm{w}^{\top} (\bm{I} - \bm{r})(\bm{c}\bm{s}^* - \min\{\bm{\mu}, \bm{c}\bm{s}^*\}) \right)}_{\eta_2} \nonumber \\
        & = \eta_1 + \eta_2 \label{eq: eta1_2}
    \end{align}
    where the time step indicator $(t)$ is dropped for simplicity. The third equal sign is obtained by introducing terms $(\bm{I} - \bm{r})\bm{c}\bm{s}^*$ and the fourth equal sign is obtained by distributing $\bm{w}^{\top}$ to each term in the bracket. $\bm{s}^*$ here denotes the signal control state determined by the generalized MP controller in Eq.~\eqref{eq: general_MP 1}.

    As indicated in Eq.~\eqref{eq: eta_delta12 1}, those movements $(i,o) \in \mathcal{M_{F}}$ do not have the term $q_{(i,o)}^{in}(s_{(i,o)}^{in}(t))$ because they are fictitious sourced links without upstream intersections. That is to say, in $\eta_2$, the corresponding values of those movements $(i,o) \in \mathcal{M_{F}}$ are 0. Thus, when we analyze $\eta_2$,  we only need to consider those movements $(i,o) \in \mathcal{M_{N}}$. 
    \begin{align}
        \eta_2 &= \mathbb{E}^{\bm{\rho}(t)} \left( \bm{w}^{\top} (\bm{I} - \bm{r})(\bm{c}\bm{s}^* - \min\{\bm{\mu}, \bm{c}\bm{s}^*\}) \right) 
        \leq \mathbb{E}^{\bm{\rho}(t)} \left( \bm{w}^{\top} \bm{I} (\bm{c}\bm{s}^* - \min\{\bm{\mu}, \bm{c}\bm{s}^*\}) \right) 
        \leq \mathbb{E}^{\bm{\rho}(t)} \left(\bm{w}^{\top} \bm{c}\bm{s}^* \right) \nonumber \\
        & \leq \mathbb{E}^{\bm{\rho}(t)}\left(\bm{w}^{\top} \bm{c} \right)
        \leq \sum_{(i,o) \in \mathcal{M}_{\mathcal{N}}} y_{(i,o)}^{max} L_i \rho_{(i,o)}^{max} c_{(i,o)} = K_2, \label{eq: ub of eta2}
    \end{align}
    where the first less-than-and-equal sign is obtained by dropping the non-positive term $- \bm{w}^{\top} \bm{r}(\bm{c}\bm{s}^* - \min\{\bm{\mu}, \bm{c}\bm{s}^*\}) $, the second less-than-and-equal sign is obtained based on the fact that $\bm{c}\bm{s}^* - \min\{\bm{\mu}, \bm{c}\bm{s}^*\} \preceq \bm{c}\bm{s}^*$, the third less-than-and-equal sign is obtained as $\bm{s}^*$ consists of binary variables, and the last less-than-and-equal sign is obtained based on Lemma \ref{lemma: upper bounds of density and flow} that $w_{(i,o)}(t) = \int_0^{L_i} y_{(i,o)}(x,t)\rho_{(i,o)}(x,t) \mathrm{d}x \leq y_{(i,o)}^{max} L_i \rho_{(i,o)}^{max}$. 

    Note that, when analyzing $\eta_1$, all movements $(i,o) \in \mathcal{M_{F \cup N}}$ are included. Based on the generalized MP controller in Eq.~\eqref{eq: general_MP 2}, we have 
    \begin{align}
        \bm{w}^\top (\bm{I}-\bm{r})\bm{c}\bm{s}^* = \max_{\bm{s} \in \bm{S}} \bm{w}^\top (\bm{I}-\bm{r})\bm{c}\bm{s} = \max_{\bar{\bm{s}} \in \bm{S}^{co}} \bm{w}^\top (\bm{I}-\bm{r})\bm{c}\bar{\bm{s}} \geq \bm{w}^\top (\bm{I}-\bm{r})\bm{c}\bar{\bm{s}}. \label{eq: mp controller for eta1}
    \end{align}
    Recall that $\bar{\bm{s}}$ denotes the long term average of signal state $\bm{s}$ and $\bm{S}^{co}$ is the convex hull of the feasible polyhedral space $\bm{S}$ of network signal states $\bm{s}$ under certain signal constraints, as defined in Definition \ref{def: admissible}. 
    Then, we have
    \begin{align}
        \eta_1 = \mathbb{E}^{\bm{\rho}(t)} \left( \bm{w}^{\top} \left( \bm{\lambda} - (\bm{I} - \bm{r})\bm{c}\bm{s}^* \right) \right) 
        \leq \mathbb{E}^{\bm{\rho}(t)} \left( \bm{w}^{\top} \left( \bm{\lambda} - (\bm{I} - \bm{r})\bm{c}\bar{\bm{s}} \right) \right)
        \leq - \epsilon \mathbb{E}^{\bm{\rho}(t)} \left( \bm{w}^{\top} \bm{1} \right)
    \end{align}
    where the first less-than-and-equal sign is obtained by substituting Eq.~\eqref{eq: mp controller for eta1} and the second less-than-and-equal sign is obtained by substituting Eq.~\eqref{eq: admissible 1}. In particular, 
    we have
    \begin{align}
        w_{(i,o)}(t) = \begin{cases}
            \int_0^{L_i} y_{(i,o)}(x,t)\rho_{(i,o)}(x,t) \mathrm{d}x \geq y_{(i,o)}^{min} z_{(i,o)}(t) > 0 \quad \text{if} \quad (i,o)\in \mathcal{M_{N}}, w_{(i,o)(t)}>0, \\
            \int_0^{L_i} y_{(i,o)}(x,t)\rho_{(i,o)}(x,t) \mathrm{d}x = z_{(i,o)}(t) = 0 \quad \text{if} \quad (i,o)\in \mathcal{M_{N}}, w_{(i,o)(t)}=0, \\
            \rho_{(i,o)}(t) = z_{(i,o)}(t) \quad \text{if} \quad (i,o)\in \mathcal{M_{F}}, \label{eq: w and rho}
        \end{cases}
    \end{align}
    where $y_{(i,o)}^{min} = \min\{y_{(i,o)}(x,t)|y_{(i,o)}(x,t) > 0, x \in [0,L_i]\} > 0$. Therefore, we can always find a positive $y^{min} \leq \min\{1, \{y_{(i,o)}^{min}\}_{(i,o) \in \mathcal{M_{N}}, w_{(i,o)}>0}\}$ that makes
    \begin{align}
        \bm{w} \succeq y^{min} \bm{z} \label{eq: w and z},
    \end{align}
    where $\bm{z}$ is the column vector of the number of vehicles $z_{(i,o)}(t)$ for movements $(i,o) \in \mathcal{M_{F \cup N}}$. Thereby, we have
    \begin{align}
        \eta_1 \leq - \epsilon \mathbb{E}^{\bm{\rho}(t)} \left( \bm{w}^{\top} \bm{1} \right) \leq  - \epsilon y^{min} \mathbb{E}^{\bm{\rho}(t)} \left( \bm{z}^{\top} \bm{1} \right) = - \epsilon' \mathbb{E} \left(\sum_{(i,o) \in \mathcal{M_{F\cup N}}} z_{(i,o)}(t) \right) \label{eq: ub of eta1}
    \end{align}
    where $\epsilon' = \epsilon y^{min}$.
    In summary, by substituting Eq.~\eqref{eq: ub of delta_0}-\eqref{eq: ub of delta_6}, Eq.~\eqref{eq: ub of eta2}, and Eq.~\eqref{eq: ub of eta1} into Eq.~\eqref{eq: sufficient conditions proof 1}, we have the Lyapunov drift 
    \begin{align}
        \mathbb{E}^{\bm{\rho}(t)} \left(\frac{\mathrm{d} V(\bm{\rho}(t))}{\mathrm{d} t}\right) & = \delta_0  - \delta_3 - \delta_4 + \delta_5 - \delta_6 + (\eta  + \delta_1 - \delta_2) = \delta_0  - \delta_3 - \delta_4 + \delta_5 - \delta_6 + (\eta_1  + \eta_2) \nonumber \\
        & \leq K - \epsilon' \mathbb{E} \left(\sum_{(i,o) \in \mathcal{M_{F\cup N}}} z_{(i,o)}(t) \right) \label{eq: ub of Lyapunov drift}
    \end{align}
    where the constant $K = K_0 + K_1 + K'_1 + K_2$. Thus, Lemma \ref{lemma: sufficient condition} is satisfied with the generalized MP controller in Eq.~\eqref{eq: general_MP 1}. According to Definition \ref{def: stability}, Theorem \ref{thm: sufficient conditions of MP} is proved. 
\end{proof}

\subsection{Stability of CV-MP in perfect environments}
In Section \ref{ssec: sufficient conditions of MP}, we have proved Theorem \ref{thm: sufficient conditions of MP} that an MP controller can stabilize the network queue if certain sufficient conditions are satisfied by its spatiotemporally varying weight on traffic flow density. In this section, we will show that the proposed CV-MP controller, which uses the real-time vehicle link travel time information as weights on traffic density, can satisfy these sufficient conditions in perfect environments.

\begin{lemma}[Upper bounded vehicle link travel time] \label{lemma: bounded travel time}
    Assume that the exogenous demand $\bm{\lambda}$ satisfies $\bm{\lambda}\in\bm{\Lambda}^{\mathrm{int}}$, where $\bm{\Lambda}$ is the admissible demand region defined in Definition~\ref{def: admissible}. Then, for any movement $(i,o)\in\mathcal{M_N}$ of a traffic network controlled by the CV-MP, there exists a finite constant $\tau_{(i,o)}^{max}$ such that for every $x\in[0,L_i]$ and all $t\ge0$, 
    \begin{align}
            0 \leq \tau_{(i,o)}(x,t) \leq \tau_{(i,o)}^{max} < \infty \label{eq: lemma travel time}
    \end{align}
    Equivalently, for every vehicle $j\in\mathcal{J}_{(i,o)}$, 
    \begin{align}
            0 \leq \tau_j(t) \leq \tau^{max} < \infty \label{eq: lemma travel time 1}
    \end{align}
    where $\tau^{max}$ is a finite constant independent of $j$ and $t$.
\end{lemma}

\begin{proof} [Proof of Lemma \ref{lemma: bounded travel time}]
    We express the travel time for a vehicle $j$ as
    \begin{align}
        \tau_j(t) = \frac{x_j(t)}{v_j} + d_j(t),
    \end{align}
    where $x_j(t)/v_j$ is the free-flow travel time and is upper bounded by $L_i/v_j$, and $d_j(t)$ is the control delay. Hence, it suffices to show that there exists a finite constant $d^{max}$ such that
    \begin{align}
        d_j(t) \leq d^{max} \quad \text{for all } j\in\mathcal{J}_{(i,o)} \text{ and } t\geq 0. \label{eq: lemma travel time 2}
    \end{align}
    Because vehicles are processed in a first-in-first-out (FIFO) manner, for any $j>1$, we have 
    $d_j(t) \le d_1(t)$
    where $d_1(t)$ is the delay of the first vehicle at the stopline. Therefore, proving
    \begin{align}
        d_1(t) \le d^{\max} \quad \forall\, t\ge0, \label{eq: lemma travel time 4}
    \end{align}
    is sufficient.
    
    We proceed by contradiction. Suppose that no finite constant $d^{max}$ exists, i.e.,
    \begin{align}
        \lim_{t\to\infty} d_1(t) = +\infty. \label{eq: lemma travel time 5}
    \end{align}
    This divergence implies one of two scenarios:
    
    \begin{enumerate}
        \item Gridlock scenario: The network becomes gridlocked, so that even during a green phase, the first queued vehicle cannot exit, leading to $d_1(t)\to\infty$. However, since $\bm{\lambda}\in\bm{\Lambda}^{\mathrm{int}}$, the network demand is strictly within the admissible limits and gridlock is precluded \citep{varaiya2013max, li2019position}.
        
        \item Perpetual red signal scenario: Alternatively, the divergence may arise because the movement’s signal remains red at all times, i.e.,
        \begin{align}
            s_{(i,o)}(t)=0 \quad \forall\, t. \label{eq: lemma travel time 6}
        \end{align}
        Under the CV-MP control policy, if the aggregate travel time (or “pressure”) for a movement diverges, then there must exist another movement $(i',o')\in\mathcal{M}_N$ with a higher pressure, ensuring that it is assigned a green phase. Formally, the control rule implies
        \begin{align}
            \sum_{j\in\mathcal{J}_{(i',o')}} \tau_j(t) - \sum_{(o',k')\in\mathcal{M}_{n'}} r_{(o',k')}(t) \sum_{j\in\mathcal{J}_{(o',k')}} \tau_j(t) > \sum_{j\in\mathcal{J}_{(i,o)}} \tau_j(t) - \sum_{(o,k)\in\mathcal{M}_{n'}} r_{(o,k)}(t) \sum_{j\in\mathcal{J}_{(o,k)}} \tau_j(t) \label{eq: lemma travel time 7}
        \end{align}
        for all $t$. 
        Since $d_1(t)$ is unbounded, Eq.~\eqref{eq: lemma travel time 7} implies 
        \begin{align}
            \lim_{t \rightarrow \infty}{\sum_{j \in \mathcal{J}_{(i',o')}} \tau_j(t) } \rightarrow \infty \Leftrightarrow \lim_{t \rightarrow \infty}{d_1(t)} \rightarrow \infty \text{ for } j=1\in \mathcal{J}_{(i',o')} \Leftrightarrow s_{(i',o')}(t) = 0 \quad \forall t. \label{eq: lemma travel time 8}
        \end{align}
    
    That is to say, the signal for movement $(i',o')$ is also always on red. Similarly, we can conclude that the signal is always red for all movement at the intersection, which leads to a contradiction with the CV-MP control rule that at least one movement should be on green.
    \end{enumerate}

    Since both scenarios lead to a contradiction—either violating the admissible demand condition or the operational principles of the CV-MP controller—we conclude that our initial assumption is false. Therefore, there exists a finite constant $d^{max}$ satisfying Eq.~\eqref{eq: lemma travel time 4}.
    This completes the proof of Lemma \ref{lemma: bounded travel time}.
\end{proof}

\begin{theorem}[Stability of CV-MP in 100\% CV environments] \label{thm: stability of CV-MP 100}
    Given the admissible demand region $\bm{\Lambda}$ in Definition \ref{def: admissible}, if the exogenous demand $\bm{\lambda} \in \bm{\Lambda^{int}}$, the proposed control policy Eq.~\eqref{eq: CV_MP}, i.e., CV-MP, can strongly stabilize the traffic network queues in 100\% CV environments where $\mathcal{J}_{(i,o)}^{cv} = \mathcal{J}_{(i,o)}$. 
\end{theorem}

\begin{proof} [Proof of Theorem \ref{thm: stability of CV-MP 100}]
     Theorem \ref{thm: stability of CV-MP 100} can be proved by showing that the spatiotemporally varying weight $\tau_{(i,o)}(t)$ where $(i,o) \in \mathcal{M_{N}}$ in the traffic density form of CV-MP, i.e., Eq.~\eqref{eq: CV_MP 2} with $\rho_{(i,o)}^{cv} = \rho_{(i,o)}$, satisfies sufficient conditions in Theorem \ref{thm: sufficient conditions of MP}.
     \begin{itemize}
         \item According to Lemma \ref{lemma: bounded travel time}, $0 \leq \tau_{(i,o)}(x,t) \leq \tau_{(i,o)}^{max} < \infty$ for any $x \in [0, L_i]$, any $t \geq 0$, and $(i,o) \in \mathcal{M_{N}}$, which satisfies \emph{Condition 1}.
         \item Regarding \emph{Condition 2}, it is obviously that $\tau_{(i,o)}(0,t) = 0$ as the vehicle link travel time $\tau_j$ in Eq.~\eqref{eq: CV_MP} is counted when $x_j > 0$. 
         \item Regarding $|\partial \tau_{(i,o)}(x,t) / \partial t|$, it is equivalent to consider $|\partial \tau_j(t) / \partial t|$. Then, we have
         \begin{align}
             \left|\frac{\partial \tau_j(t)}{\partial t}\right| &= \left|\frac{\tau_j(t+\Delta t) - \tau_j(t)}{\Delta t}\right|
             \begin{cases}
                 = t + \Delta t - t_j^0 \leq \Delta t \quad \text{if} \quad t_j^0 \in [t, t + \Delta t] \\
                 = \Delta t \quad \text{if} \quad t_j^0 < t \text{ and } \tau_j(t+\Delta t) >0 \\
                 \leq \tau_j(t) + \Delta t \quad \text{if} \quad t_j^0 < t \text{ and } \tau_j(t+\Delta t) =0 \\
             \end{cases} \nonumber \\
             & \leq \tau^{max} + \Delta t \triangleq \Dot{\tau}_{(i,o)}^{max} < \infty, \label{eq: proof of condition 3}
         \end{align}
         where $\Dot{\tau}_{(i,o)}^{max}$ is a finite constant. The first case indicates those vehicles that entered the link during $[t, t + \Delta t]$, the second case indicates those vehicles that keep on the link during $[t, t + \Delta t]$, and the third case indicates those vehicles that left the link during $[t, t + \Delta t]$ as $\tau_j(t+\Delta t) =0$. Therefore, $|\partial \tau_{(i,o)}(x,t) / \partial t|$ satisfies \emph{Condition 3}.
     \end{itemize}
     
     In summary, the spatiotemporally varying weight $\tau_{(i,o)}(t)$ in the proposed CV-MP satisfies sufficient conditions in Theorem \ref{thm: sufficient conditions of MP}, which proves Theorem \ref{thm: stability of CV-MP 100}.
\end{proof}

\begin{remark} [Encompassing existing MP controllers]
    It is not difficult to find that most existing MP controllers that are able to stabilize the network queues meet the sufficient conditions in Theorem \ref{thm: sufficient conditions of MP}. For instance,
    \begin{itemize}
        \item Q-MP lets $y_j(t) = 1$ if $x_j>0$, which makes $y_{(i,o)}(x,t) = 1$ for $x \in [0_{+}, L_{i-}]$ and $y_{(i,o)}(0,t) = 0$ and $|\partial y_{(i,o)}(x,t) / \partial t| = 0$.
        \item PW-MP lets $y_j(t) = x_j/L_i$, which makes $y_{(i,o)}(x,t) = x/L_i \leq 1$ for $x \in [0, L_i]$ and $y_{(i,o)}(0,t) = 0$ and $|\partial y_{(i,o)}(x,t) / \partial t| = 0$.
        \item D-MP lets $y_j(t) = d_j^{T_0}$ if $x_j>0$, which makes $y_{(i,o)}(x,t) \leq T_0$ for $x \in [0_{+}, L_{i-}]$ and $y_{(i,o)}(0,t) = 0$ and $|\partial y_{(i,o)}(x,t) / \partial t| \leq T_0$. 
    \end{itemize}
    Note that LC-MP is similar to PW-MP, and the difference is that LC-MP uses reinforcement learning to obtain a non-linear and time-independent $y_{(i,o)}(x,t)$, which still satisfies sufficient conditions. All of these MP controllers can be encompassed by Theorem \ref{thm: sufficient conditions of MP}.
\end{remark}

\subsection{Stability of CV-MP in imperfect environments}
In this section, we will further show that the proposed CV-MP controller can still stabilize the road network queue even in imperfect environments where CVs are heterogeneously and partially distributed. 

\begin{lemma} [Expectation of CV link travel time] \label{lemma: expectation of CV TT}
    Assuming that the probability of a vehicle $j$ on movement $(i,o) \in \mathcal{M_{F \cup N}}$ being a CV, denoted as $\beta_j = 1$, follows a Bernoulli distribution $Pr(\beta_j = 1) = \xi_{(i,o)}>0$, 
    then we have
    \begin{align}
        \mathbb{E}(w_{(i,o)}^{cv}(t)) = \xi_{(i,o)} \mathbb{E}(w_{(i,o)}(t))
    \end{align} \label{eq: expectation of CV TT}
    when $f^{sgn}(w_{(i,o)}(t)) = f^{sgn}(w_{(i,o)}^{cv}(t))$, where $f^{sgn}$ is the sign function.
\end{lemma}

\begin{proof} [Proof of Lemma \ref{lemma: expectation of CV TT}]
    With expectation, we have $\mathbb{E}(\beta_j) = \xi_{(i,o)}\times 1 + (1-\xi_{(i,o)}) \times 0 = \xi_{(i,o)}$. Then, according to Eq.~\eqref{eq: CV_MP 2.1}, when $f^{sgn}(w_{(i,o)}(t)) = f^{sgn}(w_{(i,o)}^{cv}(t))$, i.e., there are valid CV observations of movement $(i,o)$ or there is no vehicles, we have
    \begin{align}
        \mathbb{E}(w_{(i,o)}^{cv}(t)) & = \mathbb{E}\left(\sum_{j \in \mathcal{J}_{(i,o)}} \beta_j \tau_j(t)\right) = \sum_{j \in \mathcal{J}_{(i,o)}} \mathbb{E}\left(\beta_j \tau_j(t)\right) = \sum_{j \in \mathcal{J}_{(i,o)}} \mathbb{E}(\beta_j)\mathbb{E}( \tau_j(t)) = \xi_{(i,o)} \sum_{j \in \mathcal{J}_{(i,o)}} \mathbb{E}( \tau_j(t)) \nonumber \\
        & = \xi_{(i,o)} \mathbb{E}\left(\sum_{j \in \mathcal{J}_{(i,o)}}\tau_j(t)\right) = \xi_{(i,o)} \mathbb{E}(w_{(i,o)}(t)), \label{eq: proof expectation of CV TT}
    \end{align}
    which proves Lemma \ref{lemma: expectation of CV TT}.
\end{proof}

\begin{theorem}[Stability of CV-MP in imperfect CV environments] \label{thm: stability of CV-MP}
    Given the admissible demand region $\bm{\Lambda}$ in Definition \ref{def: admissible}, if the exogenous demand $\bm{\lambda} \in \bm{\Lambda^{int}}$, the proposed control policy Eq.~\eqref{eq: CV_MP}, i.e., CV-MP, can strongly stabilize the traffic network queues in heterogeneously distributed and partially CV environments with $\xi_{(i,o)}>0$ for $(i,o) \in \mathcal{M_{F \cup N}}$ and $f^{sgn}(\bm{w}(t)) = f^{sgn}(\bm{w^{cv}}(t))$ for all $t$. 
\end{theorem}
\begin{proof} [Proof of Theorem \ref{thm: stability of CV-MP}]
    From the proof of Theorem \ref{thm: sufficient conditions of MP}, we can note that in Eq.\eqref{eq: ub of Lyapunov drift}, only $\eta_1$ is determined by the signal controllers and both $\delta_0$, $\delta_3$, $\delta_4$, $\delta_5$, $\delta_6$, and $\eta_2$ are independent on the signal controllers. Thus, we only need to show that $\eta_1$ is also upper bounded under the control of the CV-MP in heterogeneously distributed and partially CV environments.

    Recall that 
    \begin{align}
        \eta_1 = \mathbb{E}^{\bm{\rho}(t)} \left( \bm{w}^{\top} \left( \bm{\lambda} - (\bm{I} - \bm{r})\bm{c}\bm{s}^* \right) \right),
    \end{align}
    where $\bm{s}^*$ is the optimal signal decision obtained by Eq.~\eqref{eq: CV_MP}. 

    Based on Lemma \ref{lemma: expectation of CV TT}, we have 
    \begin{align} \label{eq: expectation of w and w^cv}
        \mathbb{E}^{\bm{\rho}(t)}(\bm{w}) = (\bm{\xi}^{-1})^{diag} \mathbb{E}^{\bm{\rho}(t)}(\bm{w}^{cv}) 
    \end{align}
    where $\bm{w^{cv}}$ and $\bm{w}$ are the column vectors of $w_{(i,o)}^{cv}(t)$ and $w_{(i,o)}(t)$, respectively. $(\bm{\xi}^{-1})^{diag}$ is the diagonal matrix of $1/\xi_{(i,o)}$. 
    Then, we have
    \begin{align}
        \eta_1 = & \mathbb{E}^{\bm{\rho}(t)} \left( \bm{w}^{\top} \left( \bm{\lambda} - (\bm{I} - \bm{r})\bm{c}\bm{s}^* \right) \right) = (\bm{\xi}^{-1})^{diag} \mathbb{E}^{\bm{\rho}(t)} \left((\bm{w^{cv}})^{\top} \left( \bm{\lambda} - (\bm{I} - \bm{r})\bm{c}\bm{s}^* \right) \right)
    \end{align}

    Based on the proposed CV-MP controller, i.e., Eq.~\eqref{eq: CV_MP}, we have
    \begin{align}
        (\bm{w^{cv}})^\top (\bm{I}-\bm{r})\bm{c}\bm{s}^* = \max_{\bm{s} \in \bm{S}} (\bm{w^{cv}})^\top (\bm{I}-\bm{r})\bm{c}\bm{s} = \max_{\bar{\bm{s}} \in \bm{S}^{co}} (\bm{w^{cv}})^\top (\bm{I}-\bm{r})\bm{c}\bar{\bm{s}} \geq (\bm{w^{cv}})^\top (\bm{I}-\bm{r})\bm{c}\bar{\bm{s}}. \label{eq: CV-MP controller for eta1}
    \end{align}
    Then, 
    \begin{align}
        \eta_1 &= (\bm{\xi}^{-1})^{diag}\mathbb{E}^{\bm{\rho}(t)} \left((\bm{w^{cv}})^{\top} \left( \bm{\lambda} - (\bm{I} - \bm{r})\bm{c}\bm{s}^* \right) \right) \leq (\bm{\xi}^{-1})^{diag} \mathbb{E}^{\bm{\rho}(t)} \left((\bm{w^{cv}})^{\top} \left( \bm{\lambda} - (\bm{I} - \bm{r})\bm{c}\bar{\bm{s}} \right) \right) \nonumber \\
        & \leq - \epsilon \cdot  (\bm{\xi}^{-1})^{diag} \mathbb{E}^{\bm{\rho}(t)} \left( (\bm{w^{cv}})^{\top} \bm{1} \right) = - \epsilon \mathbb{E}^{\bm{\rho}(t)} \left( \bm{w}^{\top} \bm{1} \right) \leq - \epsilon' \mathbb{E} \left(\sum_{(i,o) \in \mathcal{M_{F\cup N}}} z_{(i,o)}(t) \right), \label{eq: CV-MP controller for eta1 1}
    \end{align}
    where the first less-than-and-equal sign is based on Eq.~\eqref{eq: CV-MP controller for eta1}, the second less-than-and-equal sign is based on the admissible demand region shown in Eq.~\eqref{eq: admissible 1}, and the last equal sign is obtained by substituting Eq.~\eqref{eq: expectation of w and w^cv}, and the last less-than-and-equal sign is obtained in the same way as Eq.~\eqref{eq: ub of eta1}.
    
    Combining the upper bounds of $\delta_0$, $\delta_3$, $\delta_4$, $\delta_5$, $\delta_6$, $\eta_2$ in Theorem \ref{thm: sufficient conditions of MP} and Eq.~\eqref{eq: CV-MP controller for eta1 1} , Eq.~\eqref{eq: ub of Lyapunov drift} can be obtained, which proves Theorem \ref{thm: stability of CV-MP}.
\end{proof}

\begin{corollary}[Stability of existing MP controllers in imperfect CV environments] \label{cly: existing MP imperfect CV}
    Given the admissible demand region $\bm{\Lambda}$ in Definition \ref{def: admissible}, if the exogenous demand $\bm{\lambda} \in \bm{\Lambda^{int}}$, those MP controllers encompassed in Theorem \ref{thm: sufficient conditions of MP} can also strongly stabilize the traffic network queues in heterogeneously distributed and partially CV environments with $\xi_{(i,o)}>0$ for $(i,o) \in \mathcal{M_{F \cup N}}$ and $f^{sgn}(\bm{w}(t)) = f^{sgn}(\bm{w^{cv}}(t))$ for all $t$.
\end{corollary}

\begin{proof} [Proof of Corollary \ref{cly: existing MP imperfect CV}]
    The proof of Corollary \ref{cly: existing MP imperfect CV} is the same as Theorem \ref{thm: stability of CV-MP} as these existing MP controllers can also make Eq.~\eqref{eq: expectation of w and w^cv} and~\eqref{eq: CV-MP controller for eta1} hold. 
\end{proof}

Note that, while the precondition $f^{sgn}(\bm{w}(t)) = f^{sgn}(\bm{w^{cv}}(t))$ in Theorem \ref{thm: stability of CV-MP} ensures the stability of the network by requiring that there are valid CV observations for all movements and all decision moment $t$,  it is overly strict and not necessary for stabilization. 
We acknowledge that the algorithm might still work under less-than-perfect observation conditions. Therefore, in the subsequent Theorem \ref{thm: necessary condition}, we derive a necessary condition for the stability of CV-MP in imperfect CV environments

\begin{theorem}[Necessary condition of CV-MP stability in imperfect environments] \label{thm: necessary condition}
    Let $\mathcal{M}_n^{p}$ denote the set of movements of phase $p$ at intersection $n \in \mathcal{N}$. A necessary condition of CV-MP stability in imperfect environments is that there must not exist any time $t^*$ such that 
    \begin{align}
        z_{(i,o)}^{cv}(t^*) = 0 \quad \text{and} \quad \rho_{(i,o)}(x,t^*) = \rho_{(i,o)}^{max} \quad \forall x \in [0,L_i] \quad \forall (i,o) \in \mathcal{M}_n^{p} \label{eq: necessary for CV-MP}
    \end{align}
    for all phases at all intersections.
\end{theorem}

\begin{proof} [Proof of Theorem \ref{thm: necessary condition}]
    To prove Theorem \ref{thm: necessary condition} is equivalent to proving that if there exists a time $t^*$ such that Eq.~\eqref{eq: necessary for CV-MP} holds for a phase at an intersection, the network queue will increase infinitely under the control of CV-MP.

    This is obvious. Eq.~\eqref{eq: necessary for CV-MP} is saying that for phase $p$ at intersection $n$, no CVs were observed until the traffic densities of all movements $(i,o) \in \mathcal{M}_n^p$ were at their maximum values at $t^*$. In such a case the traffic state at the incoming link that is calculated by CV-MP remains 0, i.e., $w_{(i,o)}^{cv}=0$, despite the presence of a large number of NVs in the movement $(i,o)$. In other words, we have
    \begin{align}
        s_{i,o}(t) = 0 \quad \forall t \geq t^*.
    \end{align}
    Since the traffic density of movement $(i,o)$ has reached its maximum, it has essentially been spilled over and gridlocked. Congestion will then continue to spread to the source links, destabilizing the road network.
\end{proof}

\section{Evaluation}
This section evaluates the proposed CV-MP controller at various CV environments based on a real-world corridor in Amsterdam. In particular, we select original Q-MP, PW-MP, and fully actuated control as benchmark methods for comparison. Actuated control represents a widely used real-time signal control method for urban road networks. The two MP methods (non-cyclic, as the proposed CV-MP) are then selected with the requirement that they have been demonstrated to stabilize the road network and are representative. Q-MP represents those MP controllers that use aggregated traffic metrics, and PW-MP represents those MP controllers that use spatial information of vehicles. In summary, the following methods are evaluated:
\begin{itemize}
    \item CV-MP, the proposed MP controller based on CV data, which uses the link travel time of CVs for pressure calculation.
    \item Q-MP \citep{varaiya2013max}, which uses the link length weighted vehicle counts for pressure calculation;
    \item PW-MP \citep{li2019position}, which uses position-weighted vehicle counts for pressure calculation;
    \item Fully actuated control embedded in the SUMO simulation, which switches phases when a sufficient time gap between successive vehicles is detected by loop detectors mounted 20-40 meters from the stop line.
\end{itemize}

Note that, according to Corollary \ref{cly: existing MP imperfect CV}, Q-MP and PW-MP can also theoretically stabilize the road network in imperfect CV environments. The decision step $T_0$ for all MP controllers is set at 10s. The phase transition period consists of 3s yellow time (denoted by $T_y$) without red clearance time. The saturated flow rate discount due to phase transition period is considered by multiplying $(T_0-T_y)/T_0$ when switching phases. 

\subsection{Real-world corridor}
\begin{figure}[ht!]
    \centering
    \includegraphics[width=0.8\textwidth]{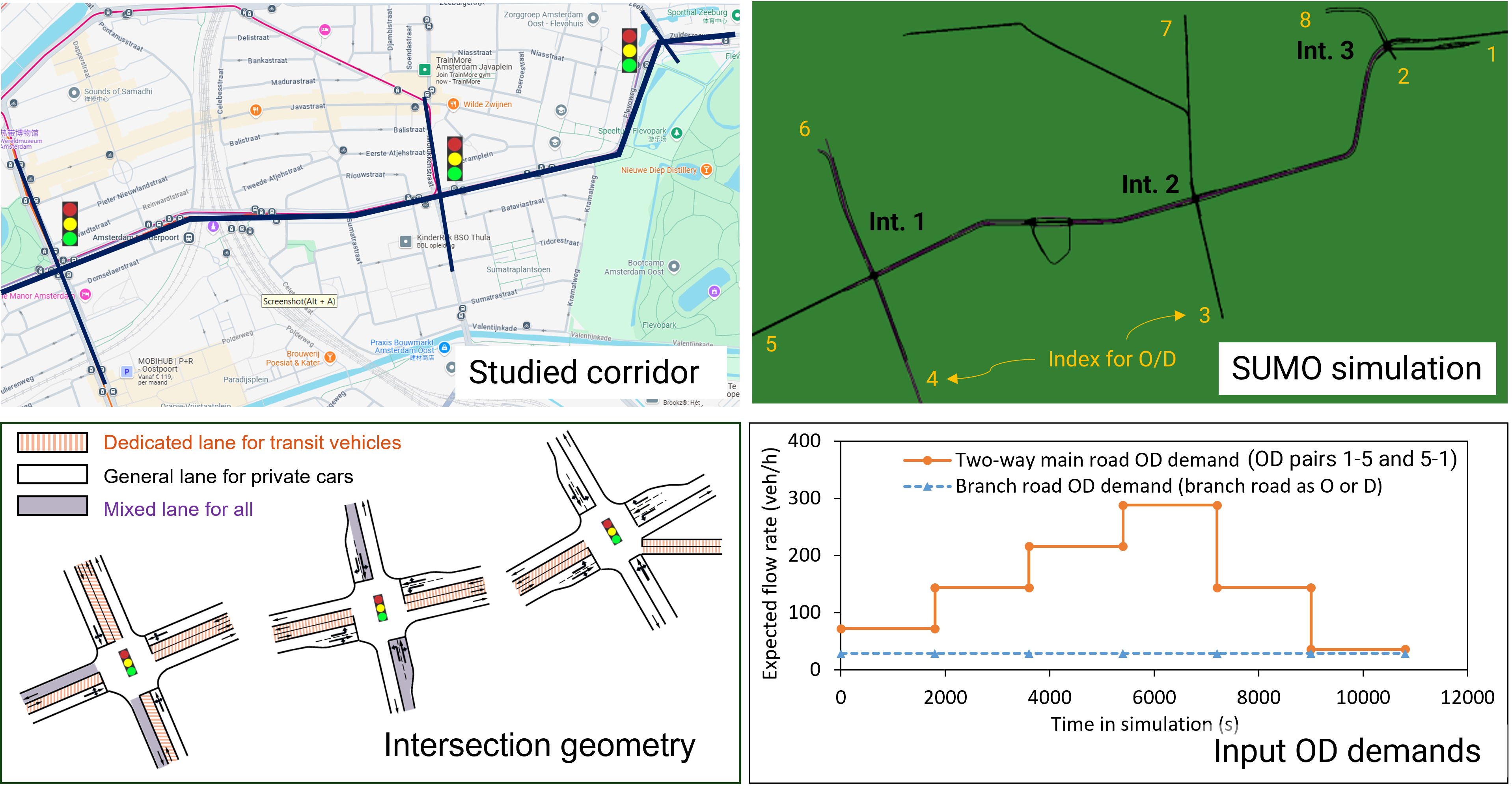}
    \caption{Simulated corridor at Amsterdam }
    \label{fig: simulation model}  
\end{figure}

A SUMO simulation network is built based on a real-world corridor in Amsterdam, as presented in Fig \ref{fig: simulation model}. The corridor consists of three signalized intersections, and the link lengths vary significantly from intersection to intersection. The longest link is over 700m, and the shortest is less than 100m. Note that only personal cars were considered in the simulation.

The input demand is based on $8 \times 7$ OD pairs, where the two-way main road OD demand (i.e., OD pairs 1-5 and 5-1) undergoes the process of increasing and then decreasing.
In total, the expected traffic demand of the corridor ranges from 1500 veh/h to 2100 veh/h during 3 hours. Note that, due to the stochastic nature of the simulation, there may be fluctuations in the total demand of 200 veh/h at different random seeds. In this study, five different random seeds are tested for each scenario.

The following metrics are used to evaluate the performance of the method:
\begin{itemize}
    \item \emph{Average vehicle delay}. The average control delay of vehicles due to signal control at intersections. Smaller delays indicate smoother traffic flow operations.
    \item \emph{Maximum vehicle counts.} The maximum number of vehicles on the studied network during the simulation. The smaller number of vehicles indicates that the method can accommodate more traffic demand. Note that this value is bounded by the number of vehicles when the road network is fully gridlocked, i.e., each lane is at maximum traffic density.
    \item \emph{Maximum queue counts.} The maximum number of vehicles that are queuing on the studied network during the simulation. Fewer queuing vehicles also indicate smoother traffic flow operations. This value is no more than the maximum number of vehicles.
    \item \emph{Maximum spillover counts.} The maximum number of vehicles planning to enter the road network but being blocked at sourced links due to spillover. The smaller number of spillover vehicles indicates that the method is effective in avoiding spillover at short lanes. In particular, the continuous increase in the number of spillover vehicles for a given demand indicates that the MP controller is not able to stabilize the network queue for that demand.
\end{itemize}

\subsection{Fully connected environments}
We first evaluate the proposed CV-MP in a perfect environment where all vehicles are connected. This scenario best reflects controller capabilities since the input data is collected perfectly. 

\begin{figure}[ht!]
    \centering
    \includegraphics[width=0.7\textwidth]{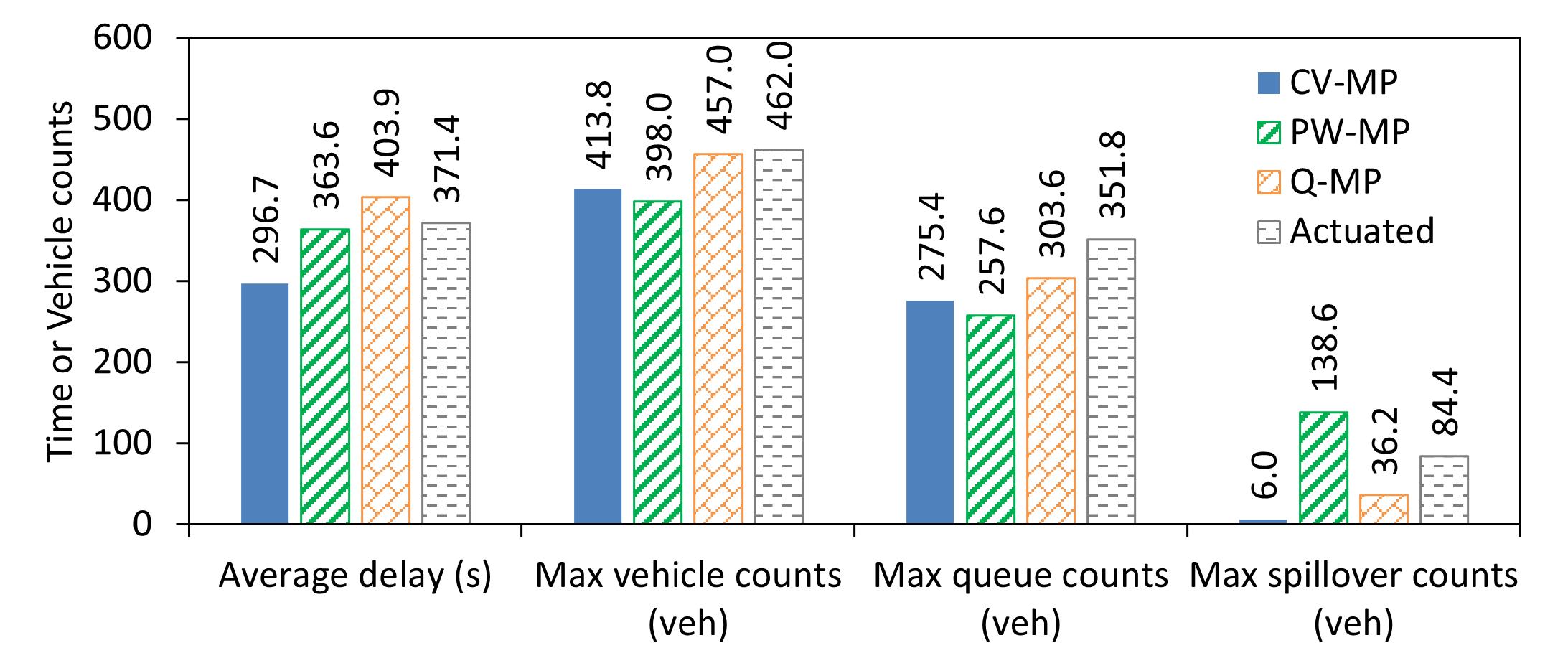}
    \caption{Overall performance in fully connected environments}
    \label{fig: overall_full}  
\end{figure}

Fig.~\ref{fig: overall_full} presents a comparative analysis of CV-MP and three benchmark methods—Q-MP, PW-MP, and actuated control—in fully connected environments, which are averaged over five different random seeds. The results indicate that CV-MP achieves the lowest average vehicle delay and effectively prevents spillover. It also performs second only to PW-MP in terms of the maximum vehicle counts and the maximum queue counts on the road network. 
While PW-MP outperforms in minimizing the maximum vehicle counts and queue counts, it suffers from significant spillover. Q-MP lags in average vehicle delay, the number of vehicles, and queuing vehicles, but demonstrates moderate spillover control, performing second only to CV-MP. Actuated control ranks lowest or second-to-last across most metrics, highlighting its limitations in our dynamic traffic demand scenarios.

\begin{figure}[ht!]
    \centering
    \includegraphics[width=0.98\textwidth]{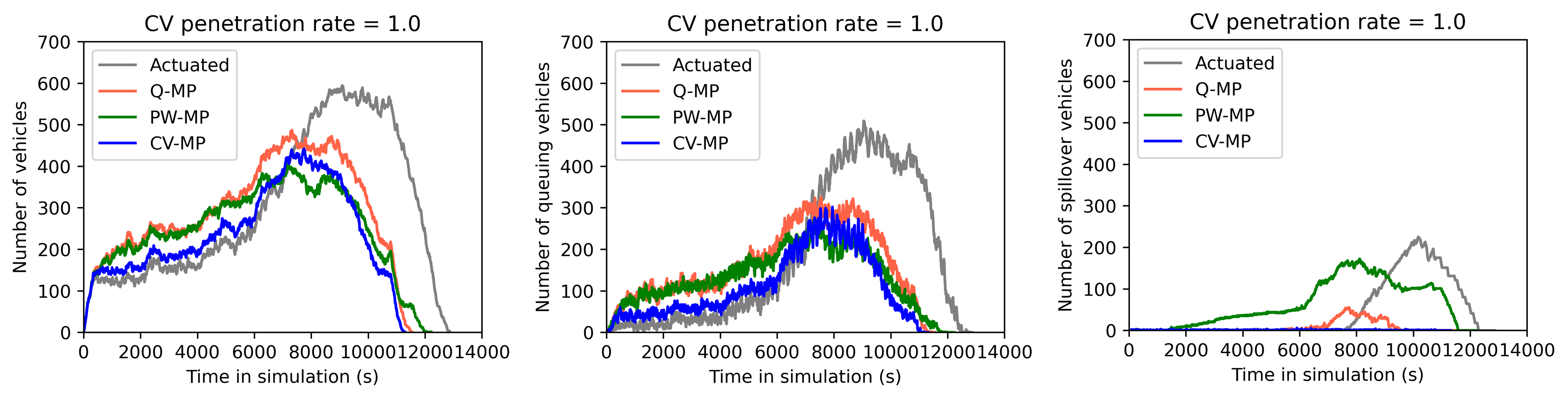}
    \caption{Detailed performance in fully connected environments}
    \label{fig: detailed_full}  
\end{figure}

\begin{figure}[ht!]
    \centering
    \includegraphics[width=0.98\textwidth]{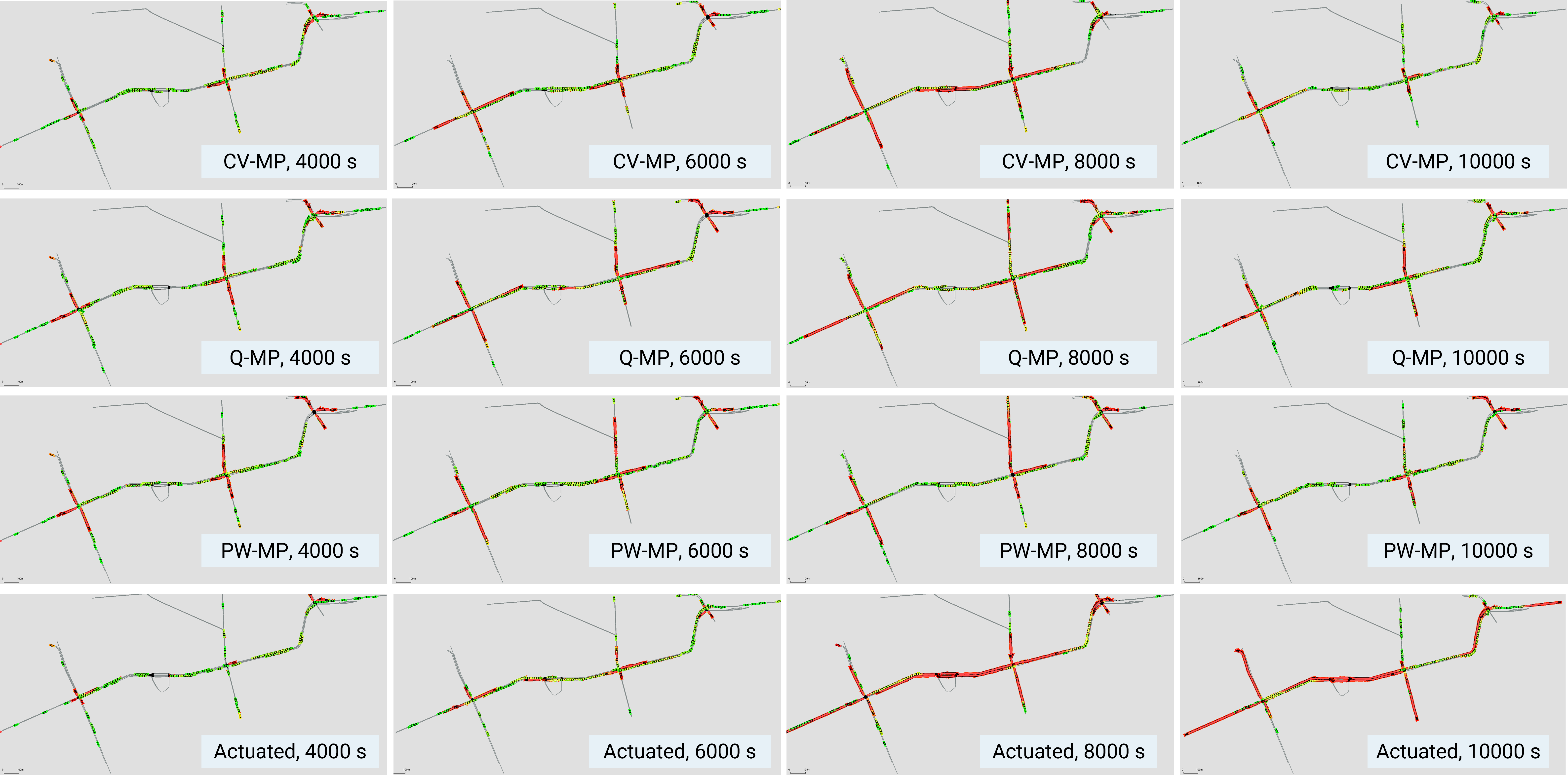}
    \caption{Representative Snapshots in fully connected environments}
    \label{fig: snapshots_full}  
\end{figure}

To investigate the underlying causes of these results, we further analyze the detailed performance of each method during the simulation, illustrated in Figures \ref{fig: detailed_full} and \ref{fig: snapshots_full}, which present the results corresponding to the random seed with higher traffic demand. In Fig.~\ref{fig: snapshots_full}, vehicle speeds are color-coded: red (stopped), yellow (half of free-flow speed), and green (free-flow speed). Basically, roads covered by red vehicles indicate the range of congestion spreading.
As shown in Fig. \ref{fig: detailed_full}, the simulation can be roughly categorized into three stages: the demand increase stage (<6000 s), the maximum demand stage (6000-9000 s), and the demand decrease stage (9000 s). Note that considering the travel time of vehicles in the road network, the real demand of the road network reflected by Fig. \ref{fig: detailed_full} will have some lag compared to the simulation input demand shown in Fig. \ref{fig: simulation model}.

At the demand increase stage, traffic demand remains relatively low. Actuated control performs best, maintaining the lowest number of vehicles, queuing vehicles, and zero spillovers. CV-MP follows closely, with slightly higher vehicle counts but still no spillover. PW-MP and Q-MP show similar trends in vehicle accumulation and queuing, though PW-MP begins to experience spillover as demand rises. Snapshots at 4000 s and 6000 s in Fig.~\ref{fig: snapshots_full} confirm these observations: actuated control results in the least congestion, followed by CV-MP, then Q-MP, then PW-MP, which shows early spillover, especially on branch roads.

As traffic demand reaches its peak during the maximum demand state, PW-MP maintains the lowest number of vehicles and queuing vehicles, but at the cost of significant spillover, particularly on branch roads. CV-MP effectively balances congestion across main and branch roads, avoiding spillover while maintaining a moderate number of vehicles. Q-MP accumulates more vehicles and queuing than CV-MP and produces only minor spillover vehicles. Actuated control fails at this stage, with a sharp rise in vehicle numbers, queuing, and widespread spillover (note that at random seeds with less demand, actuated control may perform better on avoiding spillover, leading to smaller average maximum spillover counts than PW-MP as shown in Fig.~\ref{fig: overall_full}). The 8000 s snapshot in Fig.~\ref{fig: snapshots_full} visually confirms these findings: PW-MP preserves main road performance at the expense of severe congestion on branch roads, CV-MP ensures stability by balancing congestion, Q-MP sees spillover mainly on short branch roads, and actuated control suffers extensive congestion on both main and branch roads.

The demand decrease stage reflects the efficiency of each method in dissipating congestion. CV-MP clears vehicles and queues most rapidly, restoring network stability. Q-MP follows, while PW-MP, though maintaining lower congestion on main roads, struggles with branch road spillover. Actuated control is the slowest in recovering, with persistent congestion on both main and branch roads. The 10000 s snapshot in Fig.~\ref{fig: snapshots_full} confirms these trends: CV-MP has the least congestion, Q-MP exhibits minor branch road spillover, PW-MP keeps main roads clear but retains branch congestion, and actuated control remains heavily congested.

In summary, we can conclude that:
\begin{enumerate}
    \item Actuated control is most effective at low traffic demand but deteriorates significantly as demand rises, leading to rapid congestion buildup and network destabilization under heavy traffic loads. This is mainly because in low-demand scenarios, the actuated control can switch the green phase well based on the time headway of the vehicles, while the MP control switching in a fixed decision-making step will lead to a waste of green time. With the increase of traffic demand, the actuated control will be difficult to find the best switching gap and gradually become ineffective, while the MP control can instead use all the green time for queue dissipation to achieve the maximum throughput of the road network.
    \item Q-MP performs poorly in average vehicle delay and vehicle accumulation but excels in avoiding spillover, making it a more stable alternative to PW-MP during peak demand. This is because Q-MP only considers the number of vehicles and ignores their cumulative delay as well as their spatial location, resulting in its poor performance in both average vehicle delay and maximum queue counts. However, due to the weighted link length, for links with the same number of vehicles but different lengths, the shorter link has a higher pressure value and is more likely to get a green light, resulting in relatively fewer spillovers compared to PW-MP.
    \item PW-MP prioritizes main road performance but at the cost of branch road congestion, leading to consistent spillover and network destabilization despite achieving low maximum vehicle and queuing numbers. This is due to the fact that for the same number of vehicles in the queue, longer links under PW-MP control, after considering the spatial location of vehicles, have greater pressure and receive more green lights. Our corridor scenario has significantly longer main roads than branch roads, so main roads perform significantly better than branch roads under PW-MP control.
    \item CV-MP is the most effective approach in fully connected environments, ensuring the lowest average vehicle delay, maintaining a moderate number of vehicles and queues, and preventing spillover, thereby ensuring the stability of the road network. This is mainly due to the fact that CV-MP takes into account both the spatio-temporal information of the vehicle. Even on a short branch link with fewer vehicles in the flow direction, the accumulated delay of vehicles will increase the pressure on the flow direction, allowing it to get the green light in time to avoid spillover situations.
\end{enumerate}

\subsection{Homogeneously distributed and partially connected environments} \label{ssec: homo cv}
In this section, we further evaluate the proposed CV-MP in partially CV environments with \emph{homogeneous} penetration rates across all OD pairs. Different penetration rates ranging from 0.1 to 0.9 are tested. 

\begin{figure}[ht!]
    \centering
    \includegraphics[width=0.8\textwidth]{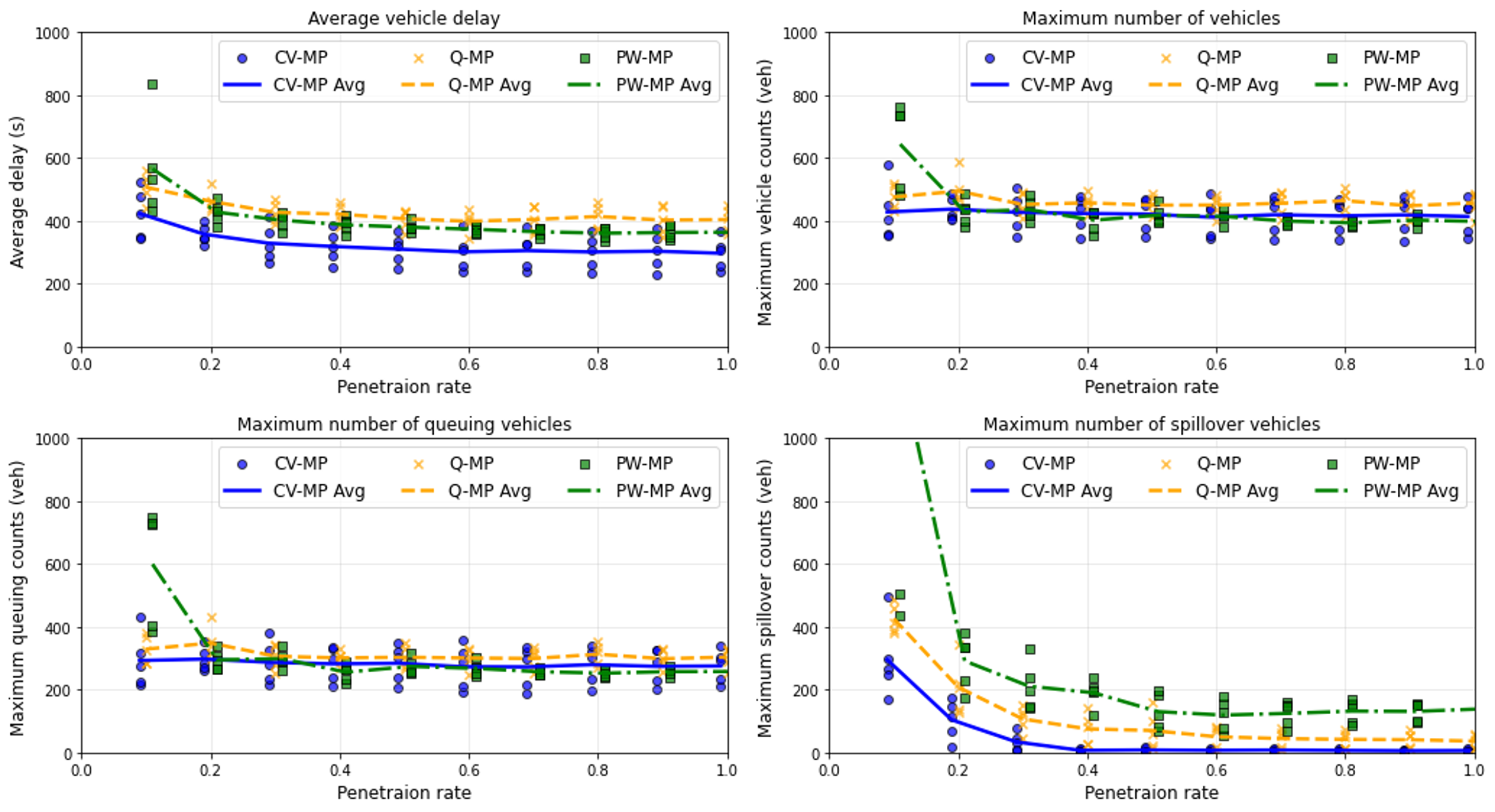}
    \caption{Overall performance in homogeneously distributed and partially connected environments}
    \label{fig: overall_homo_p}  
\end{figure}

Fig.~\ref{fig: overall_homo_p} presents the overall performance of MP controllers at different penetration rates. When the penetration rate is below 0.4, CV-MP experiences a significant reduction in average vehicle delay and spillover vehicles with the increase in penetration rate, though the marginal improvement decreases as the rate increases. Beyond a penetration rate of 0.4, both metrics stabilize. Q-MP and PW-MP exhibit similar trends, albeit at a higher penetration rate of 0.6. Regarding the maximum number of vehicles and queuing vehicles, all MP controllers show minimal sensitivity to penetration rates, except for PW-MP at a very low penetration rate of 0.1, where occasional anomalies occur.

Comparing the three MP controllers across penetration rates, their relative performance remains consistent with fully connected environments. CV-MP consistently achieves lower average vehicle delay and fewer spillover vehicles, while PW-MP minimizes the maximum number of vehicles and queuing vehicles at most penetration rates.

\begin{figure}[ht!]
    \centering
    \includegraphics[width=0.98\textwidth]{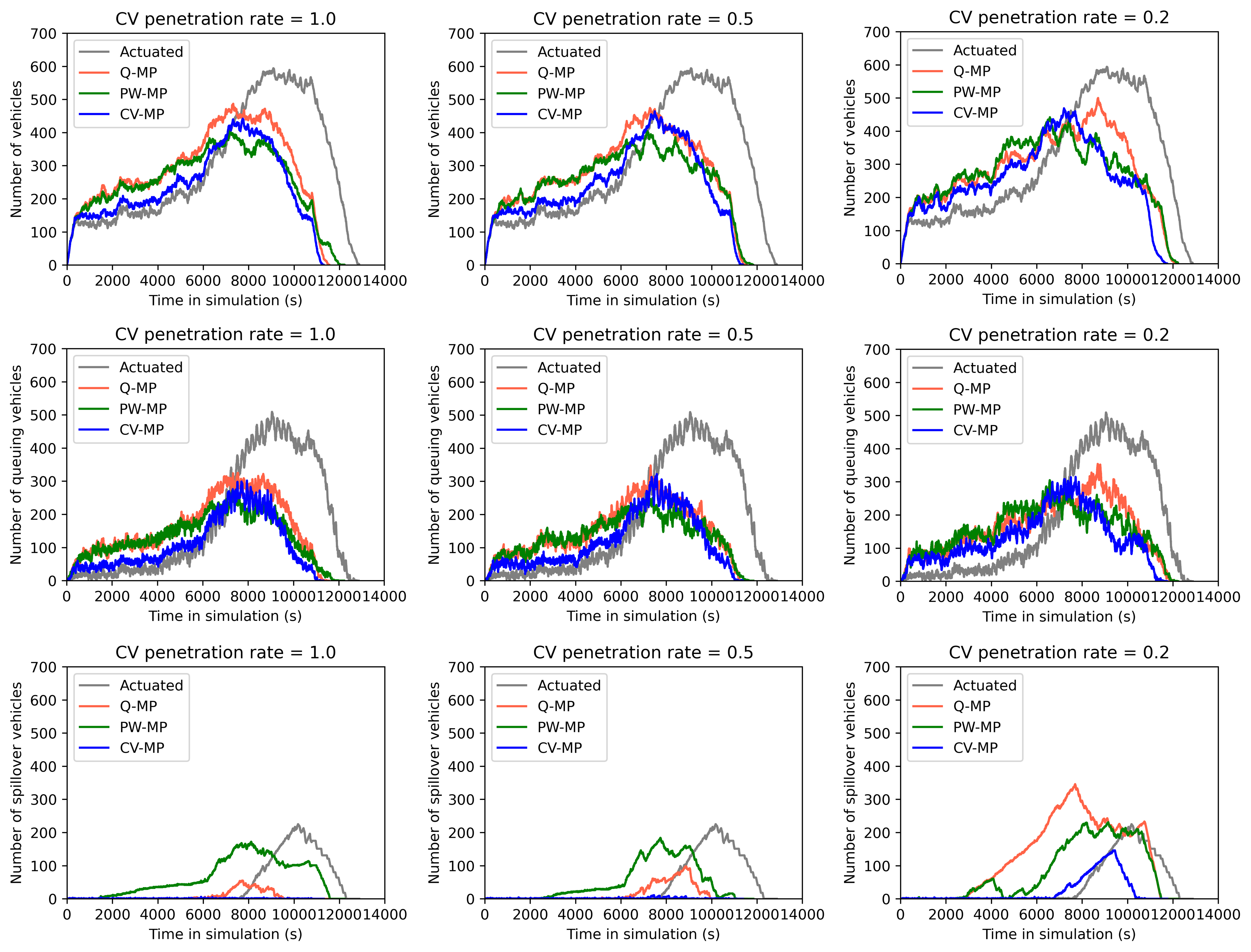}
    \caption{Detailed performance in homogeneously distributed and partially connected environments}
    \label{fig: detailed_homo_p}  
\end{figure}

Fig.~\ref{fig: detailed_homo_p} provides further insights into MP controller performance at different penetration rates. At a penetration rate of 0.5, performance closely resembles the fully connected environment. However, at a lower penetration rate of 0.2, while the maximum number of vehicles and queuing vehicles remain similar, real-time fluctuations increase significantly, suggesting heightened instability. Spillover vehicles also rise sharply for all controllers, including CV-MP, which experiences minimal spillover at higher penetration rates.

\begin{figure}[ht!]
    \centering
    \includegraphics[width=0.98\textwidth]{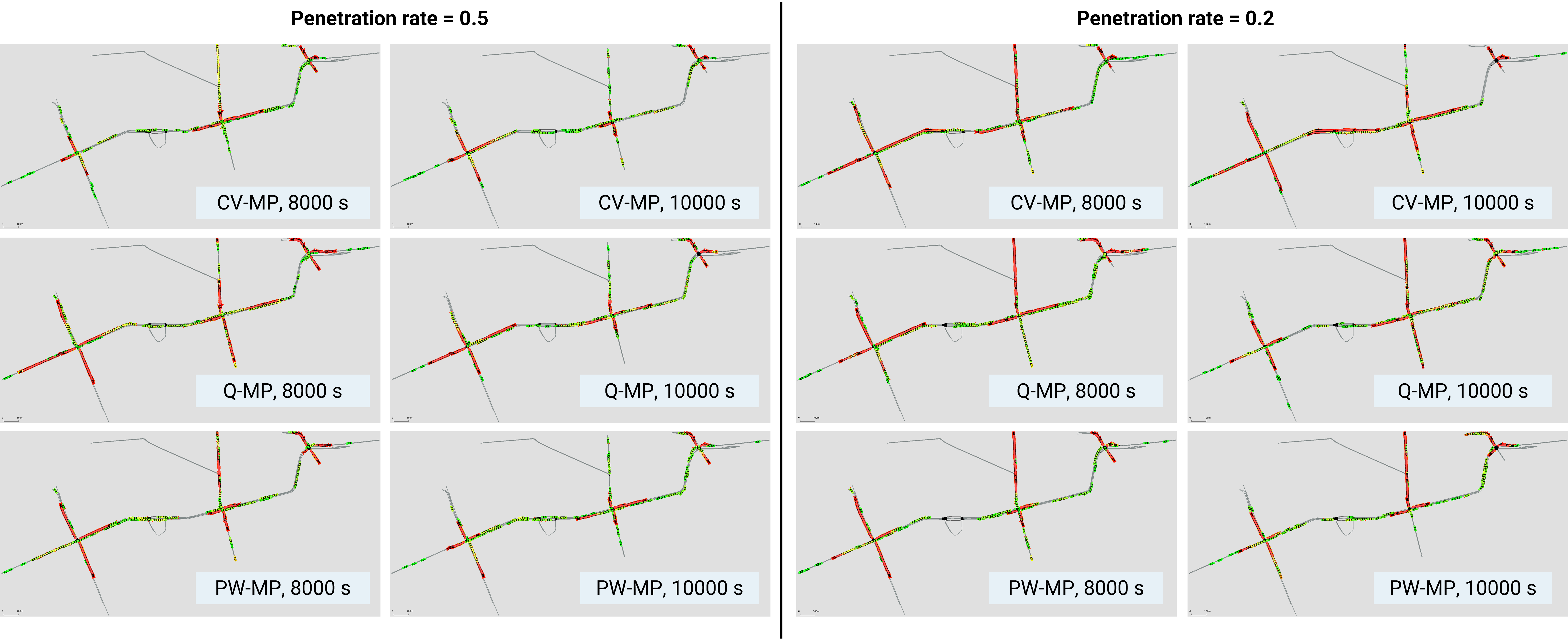}
    \caption{Representative Snapshots in homogeneously distributed and partially connected environments}
    \label{fig: snapshots_homo_p}  
\end{figure}

Snapshots in Fig.~\ref{fig: snapshots_homo_p} clarify these findings. At a penetration rate of 0.5, congestion patterns at 8000 s and 10000 s align with those observed in fully connected scenarios. However, at a 0.2 penetration rate, the random distribution of CVs results in some traffic streams lacking CV observations for extended periods. When this occurs, CV-based MP controllers fail, leading to persistent spillover and network destabilization. Snapshots at 8000 s and 10000 s reveal that branch roads experience prolonged spillover due to extended CV observation gaps.

In summary, we can conclude that:
\begin{enumerate}
    \item CV-MP consistently outperforms Q-MP and PW-MP across different penetration rates, achieving lower average vehicle delay and fewer spillover vehicles.
    \item At low penetration rates that are no more than 0.3, CV-MP experiences spillover on some branch roads, attributed to the failure of CV observations to fully satisfy the necessary condition in Theorem \ref{thm: necessary condition}—requiring valid CV observations before link spillover occurs.
\end{enumerate}

\subsection{Heterogeneously distributed and partially connected environments}\label{ssec: hetero cv}

This section evaluates the proposed CV-MP in partially CV environments with \emph{heterogeneous} penetration rates across all OD pairs. Considering that in the real world, the penetration of corridors is usually larger than that of side roads due to route choice preferences, we fix the penetration rate of the main roads of the corridor to 0.3 and set the penetration rate of side roads to 0.15, 0.2, 0.25, and 0.3, respectively, in order to test the performance of the CV-MP at different variances of penetration rates.

\begin{figure}[ht!]
    \centering
    \includegraphics[width=0.95\textwidth]{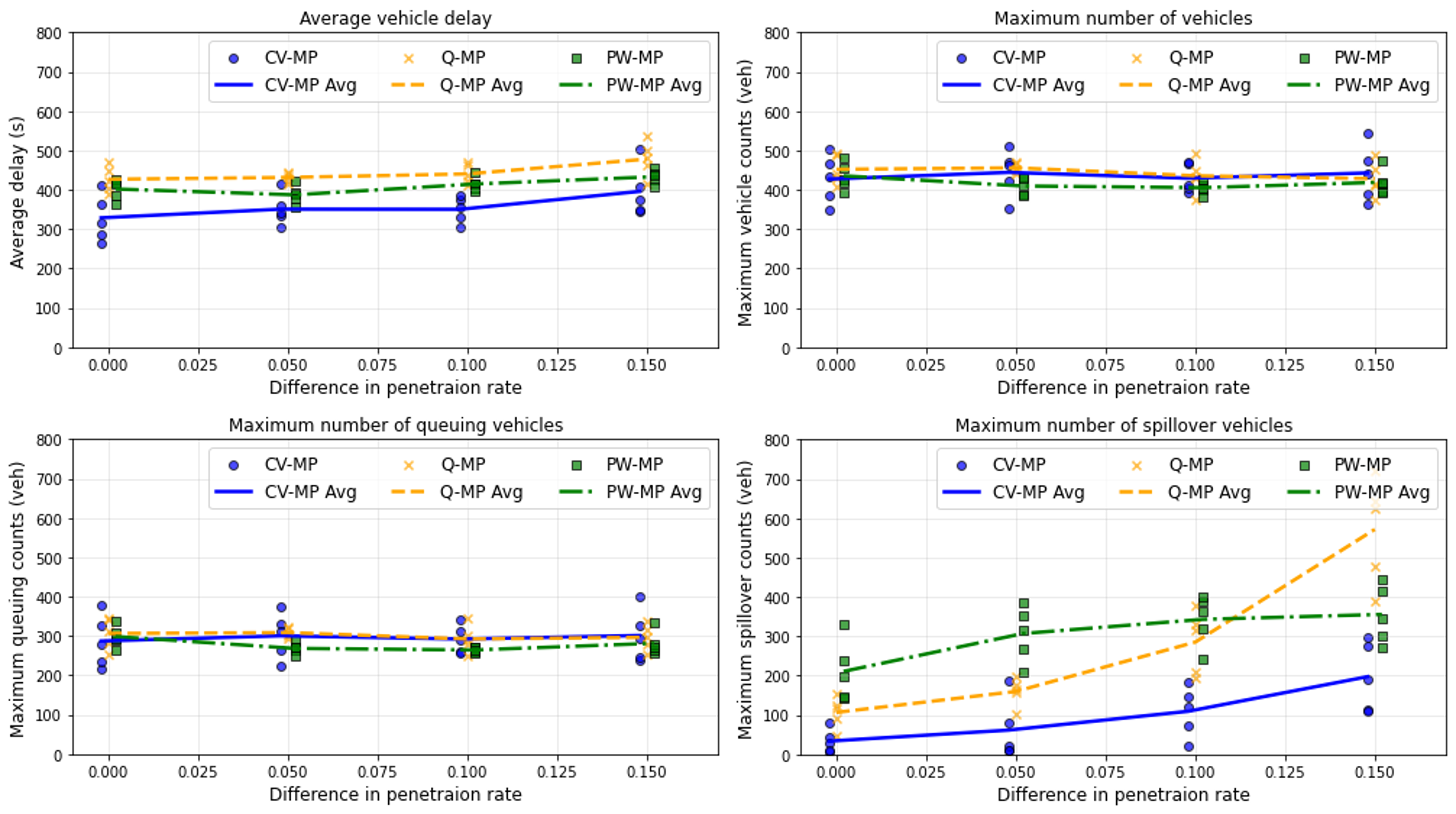}
    \caption{Overall performance in heterogeneously distributed and partially connected environments}
    \label{fig: overall_hetero_p}  
\end{figure}

The overall performance of the three MP controllers in these heterogeneous penetration environments is shown in Fig.~\ref{fig: overall_hetero_p}. The key findings include:
\begin{enumerate}
    \item CV-MP outperforms Q-MP and PW-MP across all penetration rate variances, maintaining lower average vehicle delay and fewer spillover vehicles.
    \item Higher penetration rate variance leads to increased delays and spillover vehicles for all MP controllers, mirroring the effects seen in low-penetration scenarios.
    \item For all three MP controllers, their maximum number of vehicles and queuing vehicles are insensitive to penetration rate variances. 
\end{enumerate}

The degradation in performance under greater penetration variance is attributed to the same underlying issue observed in low penetration rate environments: the necessary conditions in Theorem \ref{thm: necessary condition} are not consistently met due to insufficient CV observations. As a result, branch road signals remain in red for extended periods, exacerbating delays and spillover.

\begin{figure}[ht!]
    \centering
    \includegraphics[width=0.9\textwidth]{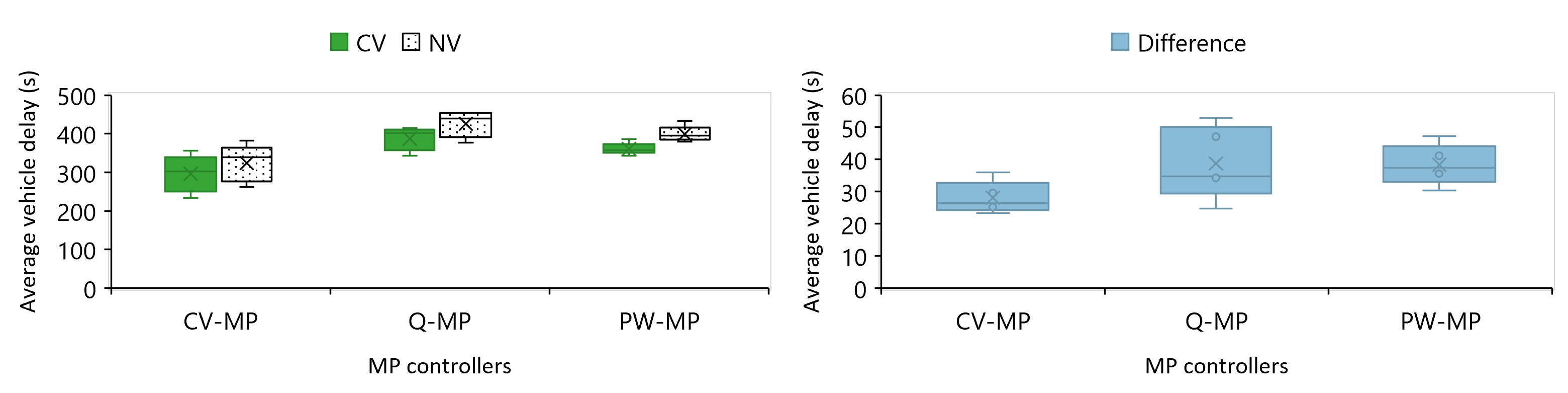}
    \caption{Average vehicle delay of CVs and NVs at a 0.5 penetration rate}
    \label{fig: cv-nv}  
\end{figure}

\subsection{Fairness between CVs and NVs}
Since CV-based traffic signal control methods leverage real-time CV data for traffic signal decisions, they may also raise fairness concerns between CVs and NVs. Therefore, this section further analyzes the fairness of these MP controls when they are based on CV data only. In order to avoid the influence of the number of vehicles on the statistic, we have performed the average delays of CVs and NVs at 0.5 CV penetration rate, i.e., when CV and NV each account for half of the traffic flow, as presented in Fig.~\ref{fig: cv-nv}.

It shows that the average CV delay is less than NV delay for all three MP controllers. However, it can be seen that the difference between the two types of vehicle delays for CV-MP is significantly smaller than that for Q-MP and PW-MP, suggesting that CV-MP can improve the fairness problem to a certain extent compared to Q-MP and PW-MP. 
This is because for MP controllers that rely on CV counts, i.e., Q-MP and PW-MP, movements with more CVs will be prioritized for the green phase, leading to a significant disparity between CV delay and NV delay. Note that PW-MP is still inherently dependent on CV counts, despite its position weighting of vehicle counts. While CV-MP, which uses travel time for pressure calculation, considers vehicle delays, it can, to some extent, reduce such disparity between CV delay and NV delay. 

\section{Conclusion and Future Work}
This study proposes a novel CV-based MP control (CV-MP) method that utilizes CV travel time data to compute pressure, fully harnessing spatiotemporal information from CVs. Unlike conventional approaches, CV-MP explicitly incorporates both the spatial distribution and temporal delays of vehicles, leading to more effective traffic regulation.
Theoretically, we rigorously prove that:
i) \emph{MP controllers with spatiotemporal-dependent weights, when satisfying certain sufficient conditions, can effectively stabilize road network queues.}
ii) \emph{CV-MP remains capable of stabilizing traffic networks, even under heterogeneously distributed and partially CV environments, provided that certain necessary conditions for CV observations are met.}

Extensive simulation validations on a real-world network in Amsterdam confirm the effectiveness of CV-MP across varying CV penetration scenarios, including fully connected, homogeneously and partially connected, and heterogeneously and partially connected cases. In all scenarios, CV-MP significantly outperforms two representative MP controllers and actuated control in reducing average vehicle delay and mitigating spillover congestion. Besides, the difference between CV delay and NV delay is smaller under CV-MP control, indicating that it is a fairer controller compared to Q-MP and PW-MP. These results demonstrate the superior traffic control performance of CV-MP with strong theoretical support for network stability.

However, several challenges remain that warrant further research.
First, the applicability of CV-MP under low CV penetration rates is constrained by the inability to meet the necessary conditions for CV-based observations. Our evaluation results in Section \ref{ssec: homo cv} and \ref{ssec: hetero cv} have shown that there is a high probability that short links in low penetration rate scenarios will fail to satisfy the necessary condition in Theorem \ref{thm: necessary condition} that at least one CV is observed before spillover. It is important to emphasize that this limitation stems from data availability rather than an inherent shortcoming of the CV-MP controller itself, as similar issues are encountered by all CV-based MP methods. A potential solution is to integrate additional sensor data to incorporate NV information, such as fixed-location detectors \citep{li2013estimating, feng2015real} or license plate recognition data \citep{zhan2015lane, tan2020fuzing, hao2024stochastic, luo2025probabilistic}, or to estimate traffic flow rates based on historical CV data \citep{zheng2017estimating, zhao2021maximum, tan2022joint}.  Incorporating such overall traffic flow information into the pressure calculation process can avoid the continuous spillover of short links due to the lack of CV observations.
Second, though CV-MP reduces the disparity in delay between CVs and NVs compared to Q-MP and PW-MP, fairness concerns still arise as such a disparity still exists. Since CV-based control strategies prioritize information-rich vehicles, they inherently benefit CVs more than NVs, leading to a potential equity issue in traffic management. Future research should explore methods to mitigate these disparities \citep{wu2018delay}.
Third, data privacy concerns emerge as CVs share real-time information for traffic control. The risk of personal data exposure necessitates robust privacy-preserving mechanisms. Prior studies, such as those by \cite{tan2024privacy, tan2025privacy}, suggest that techniques like differential privacy and secure multi-party computation can effectively protect sensitive data while maintaining the utility of shared information. Implementing such methods in CV-MP could strike a balance between operational efficiency and privacy preservation.

\printcredits

\section*{Declaration of generative AI and AI-assisted technologies in the writing process}

During the preparation of this work, the authors used ChatGPT in order to improve the language. After using this tool/service, the authors reviewed and edited the content as needed and take full responsibility for the content of the publication.

\bibliographystyle{cas-model2-names}

\bibliography{manuscript}

\begin{thebibliography}{48}
\expandafter\ifx\csname natexlab\endcsname\relax\def\natexlab#1{#1}\fi
\providecommand{\url}[1]{\texttt{#1}}
\providecommand{\href}[2]{#2}
\providecommand{\path}[1]{#1}
\providecommand{\DOIprefix}{doi:}
\providecommand{\ArXivprefix}{arXiv:}
\providecommand{\URLprefix}{URL: }
\providecommand{\Pubmedprefix}{pmid:}
\providecommand{\doi}[1]{\href{http://dx.doi.org/#1}{\path{#1}}}
\providecommand{\Pubmed}[1]{\href{pmid:#1}{\path{#1}}}
\providecommand{\bibinfo}[2]{#2}
\ifx\xfnm\relax \def\xfnm[#1]{\unskip,\space#1}\fi
\bibitem[{Ahmed et~al.(2024)Ahmed, Liu and Gayah}]{ahmed2024occ}
\bibinfo{author}{Ahmed, T.}, \bibinfo{author}{Liu, H.}, \bibinfo{author}{Gayah, V.V.}, \bibinfo{year}{2024}.
\newblock \bibinfo{title}{Occ-mp: A max-pressure framework to prioritize transit and high occupancy vehicles}.
\newblock \bibinfo{journal}{Transportation Research Part C: Emerging Technologies} \bibinfo{volume}{166}, \bibinfo{pages}{104795}.
\bibitem[{Al~Islam and Hajbabaie(2017)}]{al2017distributed}
\bibinfo{author}{Al~Islam, S.B.}, \bibinfo{author}{Hajbabaie, A.}, \bibinfo{year}{2017}.
\newblock \bibinfo{title}{Distributed coordinated signal timing optimization in connected transportation networks}.
\newblock \bibinfo{journal}{Transportation Research Part C: Emerging Technologies} \bibinfo{volume}{80}, \bibinfo{pages}{272--285}.
\bibitem[{Cao et~al.(2021)Cao, Tang, Sun and Ji}]{cao2021day}
\bibinfo{author}{Cao, Y.}, \bibinfo{author}{Tang, K.}, \bibinfo{author}{Sun, J.}, \bibinfo{author}{Ji, Y.}, \bibinfo{year}{2021}.
\newblock \bibinfo{title}{Day-to-day dynamic origin--destination flow estimation using connected vehicle trajectories and automatic vehicle identification data}.
\newblock \bibinfo{journal}{Transportation Research Part C: Emerging Technologies} \bibinfo{volume}{129}, \bibinfo{pages}{103241}.
\bibitem[{Daganzo(1994)}]{daganzo1994cell}
\bibinfo{author}{Daganzo, C.F.}, \bibinfo{year}{1994}.
\newblock \bibinfo{title}{The cell transmission model: A dynamic representation of highway traffic consistent with the hydrodynamic theory}.
\newblock \bibinfo{journal}{Transportation research part B: methodological} \bibinfo{volume}{28}, \bibinfo{pages}{269--287}.
\bibitem[{Feng et~al.(2015)Feng, Head, Khoshmagham and Zamanipour}]{feng2015real}
\bibinfo{author}{Feng, Y.}, \bibinfo{author}{Head, K.L.}, \bibinfo{author}{Khoshmagham, S.}, \bibinfo{author}{Zamanipour, M.}, \bibinfo{year}{2015}.
\newblock \bibinfo{title}{A real-time adaptive signal control in a connected vehicle environment}.
\newblock \bibinfo{journal}{Transportation Research Part C: Emerging Technologies} \bibinfo{volume}{55}, \bibinfo{pages}{460--473}.
\bibitem[{Gartner et~al.(2001)Gartner, Pooran and Andrews}]{gartner2001implementation}
\bibinfo{author}{Gartner, N.H.}, \bibinfo{author}{Pooran, F.J.}, \bibinfo{author}{Andrews, C.M.}, \bibinfo{year}{2001}.
\newblock \bibinfo{title}{Implementation of the opac adaptive control strategy in a traffic signal network}, in: \bibinfo{booktitle}{ITSC 2001. 2001 IEEE Intelligent Transportation Systems. Proceedings (Cat. No. 01TH8585)}, \bibinfo{organization}{IEEE}. pp. \bibinfo{pages}{195--200}.
\bibitem[{Goodall et~al.(2013)Goodall, Smith and Park}]{goodall2013traffic}
\bibinfo{author}{Goodall, N.J.}, \bibinfo{author}{Smith, B.L.}, \bibinfo{author}{Park, B.}, \bibinfo{year}{2013}.
\newblock \bibinfo{title}{Traffic signal control with connected vehicles}.
\newblock \bibinfo{journal}{Transportation Research Record} \bibinfo{volume}{2381}, \bibinfo{pages}{65--72}.
\bibitem[{Gregoire et~al.(2014)Gregoire, Qian, Frazzoli, De~La~Fortelle and Wongpiromsarn}]{gregoire2014capacity}
\bibinfo{author}{Gregoire, J.}, \bibinfo{author}{Qian, X.}, \bibinfo{author}{Frazzoli, E.}, \bibinfo{author}{De~La~Fortelle, A.}, \bibinfo{author}{Wongpiromsarn, T.}, \bibinfo{year}{2014}.
\newblock \bibinfo{title}{Capacity-aware backpressure traffic signal control}.
\newblock \bibinfo{journal}{IEEE Transactions on Control of Network Systems} \bibinfo{volume}{2}, \bibinfo{pages}{164--173}.
\bibitem[{Li et~al.(2013)Li, Zhou, Shladover and Skabardonis}]{li2013estimating}
\bibinfo{author}{Li, J.Q.}, \bibinfo{author}{Zhou, K.}, \bibinfo{author}{Shladover, S.E.}, \bibinfo{author}{Skabardonis, A.}, \bibinfo{year}{2013}.
\newblock \bibinfo{title}{Estimating queue length under connected vehicle technology: Using probe vehicle, loop detector, and fused data}.
\newblock \bibinfo{journal}{Transportation research record} \bibinfo{volume}{2356}, \bibinfo{pages}{17--22}.
\bibitem[{Li and Jabari(2019)}]{li2019position}
\bibinfo{author}{Li, L.}, \bibinfo{author}{Jabari, S.E.}, \bibinfo{year}{2019}.
\newblock \bibinfo{title}{Position weighted backpressure intersection control for urban networks}.
\newblock \bibinfo{journal}{Transportation Research Part B: Methodological} \bibinfo{volume}{128}, \bibinfo{pages}{435--461}.
\bibitem[{Lighthill and Whitham(1955)}]{lighthill1955kinematic}
\bibinfo{author}{Lighthill, M.J.}, \bibinfo{author}{Whitham, G.B.}, \bibinfo{year}{1955}.
\newblock \bibinfo{title}{On kinematic waves ii. a theory of traffic flow on long crowded roads}.
\newblock \bibinfo{journal}{Proceedings of the royal society of london. series a. mathematical and physical sciences} \bibinfo{volume}{229}, \bibinfo{pages}{317--345}.
\bibitem[{Lin et~al.(2012)Lin, De~Schutter, Xi and Hellendoorn}]{lin2012efficient}
\bibinfo{author}{Lin, S.}, \bibinfo{author}{De~Schutter, B.}, \bibinfo{author}{Xi, Y.}, \bibinfo{author}{Hellendoorn, H.}, \bibinfo{year}{2012}.
\newblock \bibinfo{title}{Efficient network-wide model-based predictive control for urban traffic networks}.
\newblock \bibinfo{journal}{Transportation Research Part C: Emerging Technologies} \bibinfo{volume}{24}, \bibinfo{pages}{122--140}.
\bibitem[{Liu and Gayah(2022)}]{liu2022novel}
\bibinfo{author}{Liu, H.}, \bibinfo{author}{Gayah, V.V.}, \bibinfo{year}{2022}.
\newblock \bibinfo{title}{A novel max pressure algorithm based on traffic delay}.
\newblock \bibinfo{journal}{Transportation Research Part C: Emerging Technologies} \bibinfo{volume}{143}, \bibinfo{pages}{103803}.
\bibitem[{Liu and Gayah(2023)}]{liu2023total}
\bibinfo{author}{Liu, H.}, \bibinfo{author}{Gayah, V.V.}, \bibinfo{year}{2023}.
\newblock \bibinfo{title}{Total-delay-based max pressure: a max pressure algorithm considering delay equity}.
\newblock \bibinfo{journal}{Transportation Research Record} \bibinfo{volume}{2677}, \bibinfo{pages}{324--339}.
\bibitem[{Liu and Gayah(2024)}]{liu2024n}
\bibinfo{author}{Liu, H.}, \bibinfo{author}{Gayah, V.V.}, \bibinfo{year}{2024}.
\newblock \bibinfo{title}{N-mp: A network-state-based max pressure algorithm incorporating regional perimeter control}.
\newblock \bibinfo{journal}{Transportation Research Part C: Emerging Technologies} \bibinfo{volume}{168}, \bibinfo{pages}{104725}.
\bibitem[{Liu et~al.(2024)Liu, Gayah and Levin}]{liu2024max}
\bibinfo{author}{Liu, H.}, \bibinfo{author}{Gayah, V.V.}, \bibinfo{author}{Levin, M.W.}, \bibinfo{year}{2024}.
\newblock \bibinfo{title}{A max pressure algorithm for traffic signals considering pedestrian queues}.
\newblock \bibinfo{journal}{Transportation Research Part C: Emerging Technologies} \bibinfo{volume}{169}, \bibinfo{pages}{104865}.
\bibitem[{Luo et~al.(2025)Luo, Wu, Liu, Tang and Tan}]{luo2025probabilistic}
\bibinfo{author}{Luo, L.}, \bibinfo{author}{Wu, H.}, \bibinfo{author}{Liu, J.}, \bibinfo{author}{Tang, K.}, \bibinfo{author}{Tan, C.}, \bibinfo{year}{2025}.
\newblock \bibinfo{title}{A probabilistic approach for queue length estimation using license plate recognition data: Considering overtaking in multi-lane scenarios}.
\newblock \bibinfo{journal}{Transportation Research Part C: Emerging Technologies} \bibinfo{volume}{173}, \bibinfo{pages}{105029}.
\bibitem[{Mercader et~al.(2020)Mercader, Uwayid and Haddad}]{mercader2020max}
\bibinfo{author}{Mercader, P.}, \bibinfo{author}{Uwayid, W.}, \bibinfo{author}{Haddad, J.}, \bibinfo{year}{2020}.
\newblock \bibinfo{title}{Max-pressure traffic controller based on travel times: An experimental analysis}.
\newblock \bibinfo{journal}{Transportation Research Part C: Emerging Technologies} \bibinfo{volume}{110}, \bibinfo{pages}{275--290}.
\bibitem[{Neely(2010)}]{neely2022stochastic}
\bibinfo{author}{Neely, M.}, \bibinfo{year}{2010}.
\newblock \bibinfo{title}{Stochastic network optimization with application to communication and queueing systems}.
\newblock \bibinfo{publisher}{Springer Nature}.
\bibitem[{Nishinari(2014)}]{nishinari2014traffic}
\bibinfo{author}{Nishinari, K.}, \bibinfo{year}{2014}.
\newblock \bibinfo{title}{Traffic flow dynamics: Data, models and simulation}.
\newblock \bibinfo{journal}{Physics Today} \bibinfo{volume}{67}, \bibinfo{pages}{54--54}.
\bibitem[{Rinaldi et~al.(2016)Rinaldi, Himpe and Tamp{\`e}re}]{rinaldi2016sensitivity}
\bibinfo{author}{Rinaldi, M.}, \bibinfo{author}{Himpe, W.}, \bibinfo{author}{Tamp{\`e}re, C.M.}, \bibinfo{year}{2016}.
\newblock \bibinfo{title}{A sensitivity-based approach for adaptive decomposition of anticipatory network traffic control}.
\newblock \bibinfo{journal}{Transportation Research Part C: Emerging Technologies} \bibinfo{volume}{66}, \bibinfo{pages}{150--175}.
\bibitem[{Siegel et~al.(2017)Siegel, Erb and Sarma}]{siegel2017survey}
\bibinfo{author}{Siegel, J.E.}, \bibinfo{author}{Erb, D.C.}, \bibinfo{author}{Sarma, S.E.}, \bibinfo{year}{2017}.
\newblock \bibinfo{title}{A survey of the connected vehicle landscape—architectures, enabling technologies, applications, and development areas}.
\newblock \bibinfo{journal}{IEEE Transactions on Intelligent Transportation Systems} \bibinfo{volume}{19}, \bibinfo{pages}{2391--2406}.
\bibitem[{Sims and Dobinson(1980)}]{sims1980sydney}
\bibinfo{author}{Sims, A.G.}, \bibinfo{author}{Dobinson, K.W.}, \bibinfo{year}{1980}.
\newblock \bibinfo{title}{The sydney coordinated adaptive traffic (scat) system philosophy and benefits}.
\newblock \bibinfo{journal}{IEEE Transactions on vehicular technology} \bibinfo{volume}{29}, \bibinfo{pages}{130--137}.
\bibitem[{Tan et~al.(2024a)Tan, Cao, Ban and Tang}]{tan2024connectedcfd}
\bibinfo{author}{Tan, C.}, \bibinfo{author}{Cao, Y.}, \bibinfo{author}{Ban, X.}, \bibinfo{author}{Tang, K.}, \bibinfo{year}{2024}a.
\newblock \bibinfo{title}{Connected vehicle data-driven fixed-time traffic signal control considering cyclic time-dependent vehicle arrivals based on cumulative flow diagram}.
\newblock \bibinfo{journal}{IEEE Transactions on Intelligent Transportation Systems} .
\bibitem[{Tan et~al.(2024b)Tan, Ding, Yang, Zhu and Tang}]{tan2024connected}
\bibinfo{author}{Tan, C.}, \bibinfo{author}{Ding, Y.}, \bibinfo{author}{Yang, K.}, \bibinfo{author}{Zhu, H.}, \bibinfo{author}{Tang, K.}, \bibinfo{year}{2024}b.
\newblock \bibinfo{title}{Connected vehicle data-driven robust optimization for traffic signal timing: Modeling traffic flow variability and errors}.
\newblock \bibinfo{journal}{arXiv preprint arXiv:2406.14108} .
\bibitem[{Tan et~al.(2020)Tan, Liu, Wu, Cao and Tang}]{tan2020fuzing}
\bibinfo{author}{Tan, C.}, \bibinfo{author}{Liu, L.}, \bibinfo{author}{Wu, H.}, \bibinfo{author}{Cao, Y.}, \bibinfo{author}{Tang, K.}, \bibinfo{year}{2020}.
\newblock \bibinfo{title}{Fuzing license plate recognition data and vehicle trajectory data for lane-based queue length estimation at signalized intersections}.
\newblock \bibinfo{journal}{Journal of Intelligent Transportation Systems} \bibinfo{volume}{24}, \bibinfo{pages}{449--466}.
\bibitem[{Tan and Yang(2024)}]{tan2024privacy}
\bibinfo{author}{Tan, C.}, \bibinfo{author}{Yang, K.}, \bibinfo{year}{2024}.
\newblock \bibinfo{title}{Privacy-preserving adaptive traffic signal control in a connected vehicle environment}.
\newblock \bibinfo{journal}{Transportation research part C: emerging technologies} \bibinfo{volume}{158}, \bibinfo{pages}{104453}.
\bibitem[{Tan et~al.(2021)Tan, Yao, Ban and Tang}]{tan2021cumulative}
\bibinfo{author}{Tan, C.}, \bibinfo{author}{Yao, J.}, \bibinfo{author}{Ban, X.}, \bibinfo{author}{Tang, K.}, \bibinfo{year}{2021}.
\newblock \bibinfo{title}{Cumulative flow diagram estimation and prediction based on sampled vehicle trajectories at signalized intersections}.
\newblock \bibinfo{journal}{IEEE Transactions on Intelligent Transportation Systems} \bibinfo{volume}{23}, \bibinfo{pages}{11325--11337}.
\bibitem[{Tan et~al.(2025)Tan, Yao, Tang, Liang and Yin}]{tan2025privacy}
\bibinfo{author}{Tan, C.}, \bibinfo{author}{Yao, J.}, \bibinfo{author}{Tang, K.}, \bibinfo{author}{Liang, J.}, \bibinfo{author}{Yin, G.}, \bibinfo{year}{2025}.
\newblock \bibinfo{title}{Privacy-preserving cycle-based arrival profile estimation based on cross-company connected vehicles}.
\newblock \bibinfo{journal}{IEEE Transactions on Consumer Electronics} .
\bibitem[{Tan et~al.(2022)Tan, Yao, Tang et~al.}]{tan2022joint}
\bibinfo{author}{Tan, C.}, \bibinfo{author}{Yao, J.}, \bibinfo{author}{Tang, K.}, et~al., \bibinfo{year}{2022}.
\newblock \bibinfo{title}{Joint estimation of multi-phase traffic demands at signalized intersections based on connected vehicle trajectories}.
\newblock \bibinfo{journal}{arXiv preprint arXiv:2210.10516} .
\bibitem[{Varaiya(2013a)}]{varaiya2013max}
\bibinfo{author}{Varaiya, P.}, \bibinfo{year}{2013}a.
\newblock \bibinfo{title}{Max pressure control of a network of signalized intersections}.
\newblock \bibinfo{journal}{Transportation Research Part C: Emerging Technologies} \bibinfo{volume}{36}, \bibinfo{pages}{177--195}.
\bibitem[{Varaiya(2013b)}]{varaiya2013max_1}
\bibinfo{author}{Varaiya, P.}, \bibinfo{year}{2013}b.
\newblock \bibinfo{title}{The max-pressure controller for arbitrary networks of signalized intersections}, in: \bibinfo{booktitle}{Advances in dynamic network modeling in complex transportation systems}. \bibinfo{publisher}{Springer}, pp. \bibinfo{pages}{27--66}.
\bibitem[{Wadud et~al.(2016)Wadud, MacKenzie and Leiby}]{wadud2016help}
\bibinfo{author}{Wadud, Z.}, \bibinfo{author}{MacKenzie, D.}, \bibinfo{author}{Leiby, P.}, \bibinfo{year}{2016}.
\newblock \bibinfo{title}{Help or hindrance? the travel, energy and carbon impacts of highly automated vehicles}.
\newblock \bibinfo{journal}{Transportation Research Part A: Policy and Practice} \bibinfo{volume}{86}, \bibinfo{pages}{1--18}.
\bibitem[{Wang et~al.(2020)Wang, Zhu, Hong, Wang, Tao and Wang}]{wang2020optimizing}
\bibinfo{author}{Wang, H.}, \bibinfo{author}{Zhu, M.}, \bibinfo{author}{Hong, W.}, \bibinfo{author}{Wang, C.}, \bibinfo{author}{Tao, G.}, \bibinfo{author}{Wang, Y.}, \bibinfo{year}{2020}.
\newblock \bibinfo{title}{Optimizing signal timing control for large urban traffic networks using an adaptive linear quadratic regulator control strategy}.
\newblock \bibinfo{journal}{IEEE Transactions on Intelligent Transportation Systems} \bibinfo{volume}{23}, \bibinfo{pages}{333--343}.
\bibitem[{Wang et~al.(2024)Wang, Jerome, Wang, Zhang, Shen, Kumar, Bai, Krajewski, Deneau, Jawad et~al.}]{wang2024traffic}
\bibinfo{author}{Wang, X.}, \bibinfo{author}{Jerome, Z.}, \bibinfo{author}{Wang, Z.}, \bibinfo{author}{Zhang, C.}, \bibinfo{author}{Shen, S.}, \bibinfo{author}{Kumar, V.V.}, \bibinfo{author}{Bai, F.}, \bibinfo{author}{Krajewski, P.}, \bibinfo{author}{Deneau, D.}, \bibinfo{author}{Jawad, A.}, et~al., \bibinfo{year}{2024}.
\newblock \bibinfo{title}{Traffic light optimization with low penetration rate vehicle trajectory data}.
\newblock \bibinfo{journal}{Nature communications} \bibinfo{volume}{15}, \bibinfo{pages}{1306}.
\bibitem[{Wang et~al.(2022)Wang, Yin, Feng and Liu}]{wang2022learning}
\bibinfo{author}{Wang, X.}, \bibinfo{author}{Yin, Y.}, \bibinfo{author}{Feng, Y.}, \bibinfo{author}{Liu, H.X.}, \bibinfo{year}{2022}.
\newblock \bibinfo{title}{Learning the max pressure control for urban traffic networks considering the phase switching loss}.
\newblock \bibinfo{journal}{Transportation Research Part C: Emerging Technologies} \bibinfo{volume}{140}, \bibinfo{pages}{103670}.
\bibitem[{Wu et~al.(2024)Wu, Luo, Takashi, Tang and Zhu}]{hao2024stochastic}
\bibinfo{author}{Wu, H.}, \bibinfo{author}{Luo, L.}, \bibinfo{author}{Takashi, O.}, \bibinfo{author}{Tang, K.}, \bibinfo{author}{Zhu, H.}, \bibinfo{year}{2024}.
\newblock \bibinfo{title}{Stochastic queue profile estimation using license plate recognition data}.
\newblock \bibinfo{journal}{Physica A: Statistical Mechanics and its Applications} \bibinfo{volume}{643}, \bibinfo{pages}{129790}.
\bibitem[{Wu et~al.(2018)Wu, Ghosal, Zhang and Chuah}]{wu2018delay}
\bibinfo{author}{Wu, J.}, \bibinfo{author}{Ghosal, D.}, \bibinfo{author}{Zhang, M.}, \bibinfo{author}{Chuah, C.N.}, \bibinfo{year}{2018}.
\newblock \bibinfo{title}{Delay-based traffic signal control for throughput optimality and fairness at an isolated intersection}.
\newblock \bibinfo{journal}{IEEE Transactions on Vehicular Technology} \bibinfo{volume}{67}, \bibinfo{pages}{896--909}.
\bibitem[{Wu et~al.(2020)Wu, Li, Xi and De~Schutter}]{wu2020distributed}
\bibinfo{author}{Wu, N.}, \bibinfo{author}{Li, D.}, \bibinfo{author}{Xi, Y.}, \bibinfo{author}{De~Schutter, B.}, \bibinfo{year}{2020}.
\newblock \bibinfo{title}{Distributed event-triggered model predictive control for urban traffic lights}.
\newblock \bibinfo{journal}{IEEE Transactions on Intelligent Transportation Systems} \bibinfo{volume}{22}, \bibinfo{pages}{4975--4985}.
\bibitem[{Xu et~al.(2024)Xu, Barman and Levin}]{xu2024smoothing}
\bibinfo{author}{Xu, T.}, \bibinfo{author}{Barman, S.}, \bibinfo{author}{Levin, M.W.}, \bibinfo{year}{2024}.
\newblock \bibinfo{title}{Smoothing-mp: A novel max-pressure signal control considering signal coordination to smooth traffic in urban networks}.
\newblock \bibinfo{journal}{Transportation Research Part C: Emerging Technologies} \bibinfo{volume}{166}, \bibinfo{pages}{104760}.
\bibitem[{Xu et~al.(2022)Xu, Barman, Levin, Chen and Li}]{xu2022integrating}
\bibinfo{author}{Xu, T.}, \bibinfo{author}{Barman, S.}, \bibinfo{author}{Levin, M.W.}, \bibinfo{author}{Chen, R.}, \bibinfo{author}{Li, T.}, \bibinfo{year}{2022}.
\newblock \bibinfo{title}{Integrating public transit signal priority into max-pressure signal control: Methodology and simulation study on a downtown network}.
\newblock \bibinfo{journal}{Transportation Research Part C: Emerging Technologies} \bibinfo{volume}{138}, \bibinfo{pages}{103614}.
\bibitem[{Yan et~al.(2019)Yan, He, Lin, Yu, Li and Wang}]{yan2019network}
\bibinfo{author}{Yan, H.}, \bibinfo{author}{He, F.}, \bibinfo{author}{Lin, X.}, \bibinfo{author}{Yu, J.}, \bibinfo{author}{Li, M.}, \bibinfo{author}{Wang, Y.}, \bibinfo{year}{2019}.
\newblock \bibinfo{title}{Network-level multiband signal coordination scheme based on vehicle trajectory data}.
\newblock \bibinfo{journal}{Transportation Research Part C: Emerging Technologies} \bibinfo{volume}{107}, \bibinfo{pages}{266--286}.
\bibitem[{Yao et~al.(2019)Yao, Tan and Tang}]{yao2019optimization}
\bibinfo{author}{Yao, J.}, \bibinfo{author}{Tan, C.}, \bibinfo{author}{Tang, K.}, \bibinfo{year}{2019}.
\newblock \bibinfo{title}{An optimization model for arterial coordination control based on sampled vehicle trajectories: The stream model}.
\newblock \bibinfo{journal}{Transportation research part C: emerging technologies} \bibinfo{volume}{109}, \bibinfo{pages}{211--232}.
\bibitem[{Ye et~al.(2016)Ye, Wu, Li and Mao}]{ye2016hierarchical}
\bibinfo{author}{Ye, B.L.}, \bibinfo{author}{Wu, W.}, \bibinfo{author}{Li, L.}, \bibinfo{author}{Mao, W.}, \bibinfo{year}{2016}.
\newblock \bibinfo{title}{A hierarchical model predictive control approach for signal splits optimization in large-scale urban road networks}.
\newblock \bibinfo{journal}{IEEE Transactions on Intelligent Transportation Systems} \bibinfo{volume}{17}, \bibinfo{pages}{2182--2192}.
\bibitem[{Zhan et~al.(2015)Zhan, Li and Ukkusuri}]{zhan2015lane}
\bibinfo{author}{Zhan, X.}, \bibinfo{author}{Li, R.}, \bibinfo{author}{Ukkusuri, S.V.}, \bibinfo{year}{2015}.
\newblock \bibinfo{title}{Lane-based real-time queue length estimation using license plate recognition data}.
\newblock \bibinfo{journal}{Transportation Research Part C: Emerging Technologies} \bibinfo{volume}{57}, \bibinfo{pages}{85--102}.
\bibitem[{Zhao et~al.(2021)Zhao, Wong, Zheng and Liu}]{zhao2021maximum}
\bibinfo{author}{Zhao, Y.}, \bibinfo{author}{Wong, W.}, \bibinfo{author}{Zheng, J.}, \bibinfo{author}{Liu, H.X.}, \bibinfo{year}{2021}.
\newblock \bibinfo{title}{Maximum likelihood estimation of probe vehicle penetration rates and queue length distributions from probe vehicle data}.
\newblock \bibinfo{journal}{IEEE Transactions on Intelligent Transportation Systems} \bibinfo{volume}{23}, \bibinfo{pages}{7628--7636}.
\bibitem[{Zhao et~al.(2019)Zhao, Zheng, Wong, Wang, Meng and Liu}]{zhao2019various}
\bibinfo{author}{Zhao, Y.}, \bibinfo{author}{Zheng, J.}, \bibinfo{author}{Wong, W.}, \bibinfo{author}{Wang, X.}, \bibinfo{author}{Meng, Y.}, \bibinfo{author}{Liu, H.X.}, \bibinfo{year}{2019}.
\newblock \bibinfo{title}{Various methods for queue length and traffic volume estimation using probe vehicle trajectories}.
\newblock \bibinfo{journal}{Transportation Research Part C: Emerging Technologies} \bibinfo{volume}{107}, \bibinfo{pages}{70--91}.
\bibitem[{Zheng and Liu(2017)}]{zheng2017estimating}
\bibinfo{author}{Zheng, J.}, \bibinfo{author}{Liu, H.X.}, \bibinfo{year}{2017}.
\newblock \bibinfo{title}{Estimating traffic volumes for signalized intersections using connected vehicle data}.
\newblock \bibinfo{journal}{Transportation Research Part C: Emerging Technologies} \bibinfo{volume}{79}, \bibinfo{pages}{347--362}.

\end{thebibliography}

\appendix

\section{List of important variables}  \label{Appendix list of variables}
{
\small
\begin{longtable}[htbp]{cp{12cm}}
\caption{List of important variables} \label{tab:nomenclature}\\
\hline\hline
\multicolumn{2}{l}{\textit{Section \ref{sec: pre} Preliminaries}} \\
$\mathcal{N}$ & set of all nodes of the studied network, indexed by $n$. \\
$\mathcal{M}_n$ & set of all movements for node $n$, indexed by a pair of incoming and outgoing links $(i,o)$. \\
$\mathcal{M}_{n'}$ & set of all movements for the neighboring node $n'$ of node $n$, indexed by $(o,k)$. \\
$\mathcal{F}$ & set of all fictitious source nodes. \\
$\bm{s}$ & overall signal vector composed of individual signal vector $\bm{s}_n$ for node $n$. \\
$s_{(i,o)}$ & binary variable indicating signal state for movement $(i,o)$. \\
$\bm{S}$ & feasible space of all signal states regarding signal phase constraints. \\
$\mathcal{J}_{(i,o)}, \mathcal{J}_{(i,o)}^{cv}, \mathcal{J}_{(i,o)}^{nv}$ & sets of all vehicles, CVs, and NVs of movement $(i,o)$, respectively, indexed by $j$. \\
$z_{(i,o)}, z_{(i,o)}^{cv}, z_{(i,o)}^{nv}$ & the corresponding sizes of $\mathcal{J}_{(i,o)}, \mathcal{J}_{(i,o)}^{cv}, \mathcal{J}_{(i,o)}^{nv}$. \\
$\rho_{(i,o)}$ & traffic density for the incoming link $i$ of movement $(i,o)$. \\
$q_{(i,o)}$ & flow rate for the incoming link $i$ of movement $(i,o)$. \\
$\lambda_{(i,o)}$ & exogenous demand rate for the incoming link $i$ of movement $(i,o)$. \\
$q_{(i,o)}^{out}$ & outgoing flow for the incoming link $i$ of movement $(i,o)$. \\
$s_{(i,o)}^{out}$ & downstream control state for the incoming link $i$ of movement $(i,o)$. \\
$q_{(i,o)}^{in}$ & incoming flow for the incoming link $i$ of movement $(i,o)$. \\
$s_{(i,o)}^{in}$ & upstream control state for the incoming link $i$ of movement $(i,o)$. \\
$L_i$ & link length of the incoming link $i$ of movement $(i,o)$.\\
$c_{(i,o)}$ & saturated flow rate of movement $(i,o)$ at the stopline.\\
$\mu_{(i,o)}$ & demand rate of movement $(i,o)$ at the stopline.\\
$r_{(i,o)}$ & turning rate of movement $(i,o)$. \\

\hline
\multicolumn{2}{l}{\textit{Section \ref{sec: method} Methodology}} \\
$\bm{s}^*$ & signal decision by MP controllers. \\
$y_j$ & state parameter of vehicle $j$. \\
$x_j$ & distance traveled by vehicle $j$ after entering the link. \\
$d_j$ & control delay of vehicle $j$. \\
$d_j^{T_0}$ & delay of vehicle $j$ during period $T_0$. \\
$T_0$ & time duration between two decision steps. \\
$z_{(i,o)}^{stop}$ & the number of stopped vehicles of movement $(i,o)$. \\
$\tau_j$ & normalized link travel time of vehicle $j$. \\
$t_j^0$ & the moment when vehicle $j$ entered the link. \\
$\bar{\tau}_{(i,o)}$ & average travel time of the incoming link $i$ of movement $(i,o)$. \\
$v_j$ & average free-flow speed of vehicle $j$. \\
$\beta_j$ & binary variable indicating whether the vehicle $j$ is connected. \\
$\xi_{(i,o)}$ & the penetration rate of movement $(i,o)$. \\
$\rho_{(i,o)}^{cv}$ & traffic flow density of CVs of movement $(i,o)$. \\
$\tau_{(i,o)}^{cv}$ & travel time associated weight on traffic flow density of movement $(i,o)$. \\
$w_{(i,o)}^{cv}$ & CV-based traffic state of movement $(i,o)$. \\
$\bm{w^{cv}}, \bm{r}, \bm{c}$ & vectors or matrices of $w_{(i,o)}^{cv}, r_{(i,o)}, c_{(i,o)}$ with proper size. \\
$\bm{I}$ & identical matrix with proper size. \\

\hline
\multicolumn{2}{l}{\textit{Section \ref{sec: stability} Road network stability}} \\
$V(\bm{\rho}(t))$ & Lyapunov function of traffic flow density of the road network. \\
$y_{(i,o)}$ & spatiotemporally varying weight on traffic flow density of movement $(i,o)$. \\
$\bm{\Lambda},\bm{\Lambda^{int}}$ & admissible demand region and its interior. \\
$\bar{\bm{s}}$ & long-term average of signal state $\bm{s}$. \\
$\bm{S}^{co}$ & convex hull of $\bm{S}$. \\
$w_{(i,o)}$ & traffic state of movement $(i,o)$ based on all vehicles. \\
$\bm{w}, \bm{z}$ & vectors of $w_{(i,o)}, z_{(i,o)}$ with proper size. \\
$\rho_{(i,o)}^{max}, q_{(i,o)}^{max}, y_{(i,o)}^{max}, \dot{q}_{x,(i,o)}^{max}, \dot{y}_{(i,o)}^{max}$ & upper bounds of corresponding variables. \\
$K, \epsilon', \epsilon, K_0, K_1, K_1', K_2$ & positive constants for road network stability. \\
$(\bm{\xi}^{-1})^{diag}$ & diagonal matrix of $1/\xi_{(i,o)}$. \\

\hline\hline
\end{longtable}
}

\end{document}